\documentclass[final]{siamltex}

\usepackage[a4paper,top=4.0cm,bottom=2.0cm,left=2.0cm,right=2.0cm]{geometry}

\usepackage{subcaption}
\usepackage{amsmath,amsfonts,mathrsfs,amsfonts}
\usepackage{epsfig,graphicx}
\usepackage{color}
\usepackage{bm}
\usepackage{overpic}

\usepackage[normalem]{ulem}

\newcommand{\vc}[1]{\ensuremath{\mathbf{#1}}}
\newcommand{\trp}{^{\text{\scriptsize T}} }

\newcommand{\inv}{^{-1} }
\newcommand{\sbr}[1]{\ensuremath{_{\mathrm{#1}}}}
\newcommand{\spr}[1]{\ensuremath{^{\mathrm{#1}}}}

\title{Weighted RML using ensemble-methods for data assimilation\thanks{This work of the first author was supported by the Talent Special Projects of School-level Scientific Research Programs under Guangdong Polytechnic Normal University (No.2022SDKYA025). This work of the second author was funded by the Research Council of Norway  through the Petromaks2 program (NFR project number: 295002), and by industry partners of the NORCE research cooperative research project ``Assimilating 4D Seismic Data: Big Data Into Big Models'': Equinor Energy AS, Lundin Energy Norway AS, Repsol Norge AS, Shell Global Solutions International B.V., TotalEnergies E\&P Norge AS and Wintershall Dea Norge AS.}}

\author{Yuming Ba
\thanks{School of Mathematics and Systems Science, Guangdong Polytechnic Normal University, Guangzhou 510665, China. Email:yumingba@gpnu.edu.cn. Corresponding author}
\and
Dean S. Oliver\thanks{NORCE Norwegian Research Centre, Bergen, Norway. Email: dean.oliver@norceresearch.no.}
}

\begin{document}


\maketitle


\begin{abstract}
The weighting of critical-point samples in the weighted randomized maximum likelihood method depend on the magnitude of the data mismatch at the critical points and on the Jacobian of the transformation from the prior density to the proposal density. When standard iterative ensemble smoothers are applied for data assimilation, the Jacobian is identical for all samples. If a hybrid data assimilation method is applied, however, there is the possibility that each ensemble member can have a distinct Jacobian and that the posterior density can be multimodal. In order to apply a hybrid method iterative ensemble smoother, it is necessary that a part of the transformation from the prior Gaussian random variable to the data be analytic. Examples might include analytic transformation from a latent Gaussian random variable to permeability followed by a black-box transformation from permeability to state variables in porous media flow, or a Gaussian hierarchical model for variables followed by a similar black-box transformation from permeability to state variables. 

In this paper, we investigate the application of weighting to  both types of examples.

\end{abstract}


\pagestyle{myheadings}

\thispagestyle{plain}
\markboth{Y. Ba and D. S. Oliver}{Weighted RML using ensemble-methods for data assimilation}

\section{Introduction}\label{sec:Intro}

In many geoscience applications, parameters of high-dimensional models must be estimated from a limited number of noisy data.  The data are often only indirectly and non-linearly  related to the parameters of the model. Consequently, the parameters of the model are usually underdetermined and estimation of a single set of model parameters that satisfy the data is not sufficient to characterize the solution of the inverse problem.  

Although powerful methods for quantifying uncertainty in high dimensional model spaces with Gaussian uncertainty are available \cite{martin:12} and powerful Monte Carlo methods are available for non-Gaussian low dimensional model spaces, it is still a challenge to quantify uncertainty is situations where the model dimension is large and the posterior distribution is non-Gaussian.  In that case, approximate sampling methods must typically be used.  A relatively standard approach to approximate sampling is through the minimization of a stochastic cost function. The method goes by various names including geostatistical inversing  \cite{kitanidis:95}, randomized maximum likelihood \cite{oliver:96e} and randomize-then-optimize  \cite{bardsley:14}. In this method, a realization from an approximation to the posterior distribution is generated by minimizing the weighted squared distance of the posterior sample to a realization from the prior and the squared distance between the actual data and the perturbed simulated data. The method provides exact sampling when the prior distribution is Gaussian and the relationship between data and model parameters is linear. When the posterior distribution is non-Gaussian but unimodal, it is possible to weight the realizations from minimization such that the sampling is exact  \cite{oliver:96e,oliver:17,bardsley:14,bardsley:20}.

The actual posterior landscape for distributed-parameter geoscience inverse problems is difficult to ascertain although there are known to be features of subsurface flow models that result in multimodal posterior distributions: uncertain fault displacement in a layered reservoir \cite{tavassoli:05}, uncertain rock type location in a channelized reservoir \cite{zhang:03b}, layered non-communicating reservoir flow with independent uncertain properties \cite{oliver:11b}, and non-Gaussian prior distributions for log-permeability \cite{oliver:18a}. For relatively simple transient single-phase flow problems, in which the prior distribution of log-permeability is multivariate normal, the posterior distribution appears to be multimodal in low-dimensional subspaces, but appears to be characterized by curved ridges in higher dimensional subspaces \cite{oliver:11}. 

Sampling the posterior distribution correctly for subsurface flow models is difficult when there are hundreds of thousands to millions of uncertain parameters whose magnitudes must be inferred from the data. The gold standard for sampling from Bayesian posteriors is usually considered to be Markov chain Monte Carlo (MCMC) methods.   Because of the high computational cost of evaluating the likelihood function for subsurface flow models,  MCMC is seldom used for subsurface models, however.   For a 2D single phase porous flow problem in which only the permeability field was uncertain, an MCMC method with transition proposals required hundreds of thousands of iterations to generate a useful number of independent samples \cite{oliver:97}.  For a porous flow model with a more complex posterior but only three uncertain parameters,   a population MCMC approach showed good mixing properties and  gave good results, but  at substantial cost \cite{mohamed:12}.  In order to reduce the cost of the likelihood evaluation, Maschio and Schiozer \cite{maschio:14} replaced the flow simulator by proxy models generated by an artificial neural network.  An iterative procedure combining MCMC sampling and ANN training was applied  to a reservoir model with 16 uncertain attributes.

Iterative ensemble smoothers, on the other hand, have been remarkably successful at history matching large amounts of data into reservoir models with hundreds of thousands of parameters. Based loosely on the ensemble Kalman filter \cite{evensen:94} which is routinely used for numerical weather prediction,  iterative ensemble smoothers  use stochastic gradient approximations for minimization, so that solutions of the adjoint system are not necessary.  The downside of this is that the methodology must approximate the cost function as a quadratic surface. The method is not well suited to arbitrary posterior landscapes. 

When the posteriori pdf has multiple modes of any type,  minimisation-based simulation methods  will almost certainly  sample occasionally from local minima of the cost function that contribute very little to the probability mass in the posterior.  These samples are usually, but not always, characterized by large data mismatch after minimization. In the case of an exceptionally large data mismatch, it is common practice to omit the offending realization when computing mean forecasts and uncertainty quantification.  It is much more difficult to decide how to treat realizations with intermediate data mismatches, or realizations in general when the unweighted distribution is known to be only approximate.  Importance weighting of realizations to correct for the approximate sampling  is the principled approach to uncertainty quantification in these cases.

Van Leeuwen et al.\ \cite{vanLeeuwen:19} identify four approaches to reducing the variance in weights for particle filters. The minimization approach (RML) used in this paper can be considered to belong to the category in which particles are pushed from the prior into regions of high posterior probability density.  For inverse problems with Gaussian priors on model parameters and linear observation operators, after minimization the particles are distributed as the target distribution so weighting is not required. When the posterior is multimodal, however, the problem of weighting can be relatively complex as each particle in the prior potentially maps to multiple particles in the proposal density, each with a different weight. Ba et al.\ \cite{ba:22} computed low-rank approximations to the particle weights for a single-phase porous media flow problem using the adjoint system for the flow simulator, but adjoint systems are not always available and the cost can be prohibitive. In this paper we show how approximate weights can be easily computed when an ensemble Kalman-based approach is used to solve an inverse problem.

\subsection{Example applications} \label{sec:intro_example_applications}

The weighting of critical-point samples in the randomized maximum likelihood method depends on the magnitude of the data mismatch at the critical points and on the Jacobian of the transformation from the prior density to the proposal density. When standard iterative ensemble smoothers are applied for data assimilation, the Jacobian is identical for all samples. If a hybrid data assimilation method\footnote{Hybrid method can refer to many different approaches. Here we refer to approaches that use gradients that are computed partially from the ensemble and partially by direct differentiation.} is applied, however, there is the possibility for each ensemble member to have a distinct Jacobian and for the posterior distribution of particles to be multimodal. In order to apply a hybrid method iterative ensemble smoother, it is necessary that a part of the transformation from the prior Gaussian random variable to the data be analytic. Examples might include transformation from a latent Gaussian random variable to permeability followed by a system of partial differential equations mapping permeability to state variables in porous media flow, or a Gaussian hierarchical model for variables followed by a similar  transformation from permeability to state variables
\begin{equation} \label{eq:hy_grad_M}
G = (\nabla_x(g^T))^T =  G_m (\nabla_x( m^T))^T= G_m  M_x.
\end{equation}

\paragraph{Hierarchical Gaussian}

For a  hierarchical Gaussian model in which hyper-parameters, $\theta$, of the prior model covariance such as the principal ranges and the orientation of the anisotropy are uncertain, we might use  the non-centered parameterization \cite{papaspiliopoulos:07} to express the relationship between the observable Gaussian variable $m$ and the model parameters $z$ and $\theta$  as
\begin{equation*}
m= m_{pr}+ L(\theta) z,
\end{equation*}
where the model covariance matrix $C_m= L L^T$.  In this application, the sensitivity of the observable variable $m$ to the latent variables $z$ and $\theta$ is non-local,
\begin{equation*} 
M_x  = \begin{bmatrix} L &    (\nabla_\theta  L) z \end{bmatrix} .
\end{equation*}
In a hybrid IES, the sensitivity of production data to permeability and porosity, $G_m$,  would be estimated using the ensemble of predicted data and the ensemble of model perturbations. 

\paragraph{Transformation of permeability}

In some applications of data assimilation to subsurface characterization, it is desirable to generate prior realizations of the permeability field with non-Gaussian structure, i.e.\  continuous channel-like features of high permeability embedded in a low permeability background. A property field with these  characteristics can be obtained by applying a  nonlinear transformation  to a correlated Gaussian field, i.e. $m = f(x)$ where $x\sim {\rm N}(x^\text{pr}, C_x)$.  Unlike the hierarchical example, in which the sensitivity matrix $M_x$ had dimensions $N_m \times N_m$ and was potentially full, the sensitivity in this application will generally be diagonal.

\paragraph{Composite observation operators}

For some types of subsurface data assimilation  problems, the observation operator might be separable into two parts, one of which can be treated analytically, and the other might be a complex relationship, determined by the solution of a partial differential equation.  An example is the observation of acoustic impedance in a seismic survey. The impedance, $Z$ is related to the state of the reservoir (i.e.\ the pressure and saturation) and other reservoir properties, $\theta$, through a rock physics model, which may have several uncertain parameters. The state of the reservoir $u$ is a function of permeability and porosity which we denote as $u(x)$. The composite relationship is written loosely as
\begin{equation*} 
Z(x,\theta) = Z(u(x),\theta).
\end{equation*}
The sensitivity of impedance to the permeability and porosity can be computed as
\begin{equation} \label{eq:hy_grad_Z}
G = (\nabla_x(Z^T))^T =  (\nabla_{u,\theta} ( Z^T))^T (\nabla_x( u^T))^T= Z_{u,\theta}  u_x 
\end{equation}
in which case, the sensitivity of impedance to the state variables and parameters of the rock physics model, $Z_{u,\theta}$, can be computed analytically, and the sensitivity of the state variables to permeability and porosity can be computed stochastically as in an iterative ensemble smoother.

\paragraph{Novelty}

In this paper, we develop an importance weighting approach to the problem in which the relationship between observations and model parameters is sufficiently nonlinear that the posterior distribution for model parameters is multimodal.  Although importance weighting in particle filters is a standard approach for dealing with nonlinearity in small data assimilation problems, we apply it to problems with relatively large numbers of model parameters and data. We show that this can be done in a hybrid iterative ensemble smoother approach for which gradients required for minimization are computed using a combination of analytical and stochastic gradients.  The method is applied to a two-phase porous media flow problem with multimodal posterior probability density function. In order to be useful for large problems, we improve an earlier approach through the use of circulant embedding of the covariance matrix to allow matrix multiplication in large models. Finally, we demonstrate that the weights computed using a hybrid  or ensemble Kalman-like approach are noisy approximations of the true weights and  that denoising the  weights improve model predictability.


\section{Methodology}

Consider the following generic forward model for prediction of $u$ given $x$,
\[
{\mathcal B} (x,u)=0, \quad \text{in}\quad \Omega,
\]
which, for example, could be a system of  partial differential equations (PDEs) characterizing a physical problem. In the parameter estimation problem, the task is to quantify the unknown parameter $x$ given some limited observations of $u$ on parts of the domain $\Omega$. The relationship between model parameters and observations is given by the widely used model:
\[
\bm d^o=g(m(x))+\bm\epsilon,
\]
where $g(\cdot)$ is the generic observation operator and $m(\cdot)$ is the model operator mapping the unknown $x$ to the space of intermediate variable, such as the hierarchical model, transformation of permeability and composite observation,  including the three cases of Sec.~\ref{sec:intro_example_applications}. In a finite dimensional parameter space, $\bm x\in\mathbb R^{N_x}$, $\bm m(\bm x)\in\mathbb R^{N_m}$ and $\bm d^o\in\mathbb R^{N_d}$. Assume that $\bm\epsilon \in\mathbb R^{N_d}$ is independent of $\bm x$ and $\bm\epsilon\sim {\rm N}(\vc 0,\vc C_d)$. Given a prior Gaussian  distribution ${\rm N}({\bm x}^\text{pr}, \vc C_x)$, we expect to generate samples $\bm x_i$, $i=1,\cdots,N_e$, from the posterior distribution
\[
\pi_X(\bm x|\bm d^o)=\frac{\pi_{XD}(\bm d^o|\bm x)}{\pi_D(\bm d^o)}\propto\exp(-O(\bm x))
\]
with the negative log posterior function
\[
O(\bm x)=\frac{1}{2}(\bm x-{\bm x}^\text{pr})^T\bm C_x\inv(\bm x-{\bm x}^\text{pr})+\frac{1}{2}(g(\bm m(\bm x))-\bm d^o)^T\bm C_d\inv(g(\bm m(\bm x))-\bm d^o).
\]
In general, the normalisation constant $\pi_D(\bm d^o)$ is unknown, but independent of $\bm x$. For simplicity, $\bm m(\bm x)$ is denoted as $\bm m$.

In this paper, we  apply ensemble Kalman-like approximations to the randomized maximum likelihood method \cite{kitanidis:95,oliver:96e,chen:12} for data assimilation.
The randomized maximum likelihood (RML) method draws samples $(\bm x'_i, \bm\delta'_i)$ from the Gaussian distribution
\[
q_{X'\Delta'}(\bm x',\bm\delta')=\frac{1}{(2\pi)^{\frac{N_xN_d}{2}}|\bm C_x|^{1/2}|\bm C_d|^{1/2}}\exp\Big(-\frac{1}{2}(\bm x'-{\bm x}^\text{pr})^T\bm C_x\inv(\bm x'-{\bm x}^\text{pr})-\frac{1}{2}(\bm \delta'-\bm d^o)^T\bm C_x\inv(\bm \delta'-\bm d^o)\Big)
\]
for given ${\bm x}^\text{pr}$ and $\bm d^o$. The $i$th approximate  posterior sample is then generated by computing the critical points of the cost functional
\begin{equation}\label{eq:sto_mini}
O_i(\bm x)=\frac{1}{2}(\bm x-\bm x_i')^T\bm C_x\inv(\bm x-\bm x_i')+\frac{1}{2}(g(\bm m)-\bm\delta'_i)^T\bm C_d\inv(g(\bm m)-\bm\delta'_i).
\end{equation}
The critical points are obtained by solving $\nabla_xO_i(\bm x)=0$ for $\bm x$. In general, the maxima and stationary points contribute little and  Levenberg-Marquardt with a Gauss-Newton approximation of the Hessian is used for the minimization.  The $i$th increment in the iteration is written as
\begin{equation}\label{eq:rml}
\delta \bm x_l=\frac{\bm x'_i-\bm x_l}{1+\lambda_l}-\bm C_x\bm G_l^T\Big[(1+\lambda_l)\bm C_d+\bm G_l\bm C_x\bm G_l^T\Big]\inv 
\bigg[\Big(g(\bm m_l)-\bm\delta_i'\Big)-\frac{\bm G_l(\bm x_l-\bm x'_i)}{1+\lambda_l}\bigg],
\end{equation}
where $\bm G_l=(\nabla_{x_l}(g^T))^T$ and $\lambda_l$ is the Levenberg-Marquardt regularization parameter for the $\ell$th iteration.

The ensemble-Kalman approximation of  RML is asymptotically exact for Gauss-linear  data assimilation problems and adopts an average sensitivity computed from the ensemble samples to approximate the downhill direction \cite{chen:12}, which results in an inaccurate sensitivity when the problem is highly nonlinear. 
To improve the accuracy of the sensitivity matrix for individual realizations, the hybrid ensemble-method is introduced. Through proper forms such as Eq.~\eqref{eq:hy_grad_M} an Eq.~\eqref{eq:hy_grad_Z} in hybrid ensemble-method, some derivatives are computed analytically, and others are approximated from the ensemble. 
Consequently, instead of a single common gain matrix applied to all realizations, 
the gain matrix of each sample in hybrid ensemble-methods is different. In a na{\"i}ve implementation, the computation cost will be very high for large models. We take advantage of the block-circulant structure of the prior model covariance matrix or the square root of it to reduce the cost substantially, applying circulant embedding for fast multiplication of Toeplitz matrices.

For posterior distributions with multiple modes, the approximate samples from  minimisation-based simulation methods will almost certainly converge to local minima, some of which contribute very little to the probability mass in the posterior. These samples may result in large data mismatch after minimization. Importance weights of approximate samples from the proposal distribution are used to correct the sampling. To compute the importance weights, it is necessary to compute the proposal distribution for RML samples.  Solving $\nabla_x O(\bm x)=0$ leads to a map from $(\bm x, \bm\delta)$  to $(\bm x', \bm\delta')$ in Ba et al.\ \cite{ba:22},
\begin{equation}\label{eq:rml_map}
\left\{\begin{aligned}
&\bm x'=\bm x+\bm C_x\bm G^T\bm C_d\inv\big(g(\bm m)-\bm\delta\big)\\
&\bm\delta'=\bm\delta.
\end{aligned}\right.
\end{equation}
Based on the map of Eq.~\eqref{eq:rml_map} and the original notation, the transformed distribution is given by
\begin{equation}\label{eq:prop}
p_{X\Delta}(\bm x,\bm\delta):=n(\bm x')\inv q_{X'\Delta'}(\bm x',\bm\delta')J(\bm x, \bm \delta)=n(\bm x')\inv q_{X'}\Big(\bm x+\bm C_x\bm G^T\bm C_d\inv\big(g(\bm m)-\bm\delta\big)\Big)q_{\Delta'}(\bm\delta)J(\bm x, \bm \delta),
\end{equation}
where $n(\bm x')$ is the total number of critical points of Eq.~\eqref{eq:sto_mini} and $J(\bm x,\bm\delta)$ denotes the Jacobian determinant associated with the map $(\bm x,\bm\delta)\rightarrow (\bm x',\bm\delta')$. In the following, we assume that the map is locally invertible, i.e., $J\neq 0$ everywhere. The form of $J(\bm x,\bm\delta)$ is provided by
\[
J(\bm x,\bm\delta)=\bigg|I+\mathcal{D}\Big(\bm C_x\bm G^T\bm C_d\inv\big(g(\bm m)-\bm\delta\big)\Big)\bigg|.
\]

When importance sampling is implemented for highly nonlinear problems, the variance in the log-weights will generally not be small. This is due to the fact that the RML proposal density is not identical to the target density and the  ensemble samples are only approximations of the samples that would be obtained from exact computation of minima. Because of the approximations, the actual spread in computed importance weights will be larger than it should be. Denoising of importance weights has been shown to be effective at improving the weights \cite{akyildiz:17}. For the ensemble-methods based on RML, the denoising will be done when the variance of weights is large.


\subsection{Data assimilation based on ensemble methods}

In practice, iterative ensemble smoothers are often an effective approach for solving large-scale geoscience inverse problems. These  methods are based on the Kalman filter \cite{evensen:94}, which uses a low-rank approximation of the covariance matrix to replace the full covariance and avoids the need to compute adjoints of the objective functions as might be required in an extended Kalman filter. To improve the efficiency of updating the unknown parameters, a `smoother' method using all the data simultaneously is generally used for parameter estimation problems. However, most parameter estimation problems are nonlinear and a single update in which all data are simultaneously assimilated is not enough. For history matching, iteration is a necessary of a smoother application. Iterative ensemble smoothers (IES) and their variants include two general approaches: multiple data assimilation (MDA) \cite{reich:11,emerick:13} and IES based on randomized maximum likelihood (RML) \cite{chen:12}. The iterative ensemble smoother form of RML uses an average sensitivity to approximate the Hessian matrix. For the strongly nonlinear problems, the ensemble average sensitivity will provide a poor approximation of the local sensitivity. To partially rectify this problem,  a hybrid RML-IES method has been proposed to improve the estimate of the local  sensitivity;  some  gradients are computed analytically and others are approximated from the ensemble \cite{oliver:22a}.

\subsubsection{Iterative ensemble smoother}

For the RML method, the computation of the gradient of the objective function with respect to the parameters is necessary. In many high-dimensional problems, the computation of derivatives is difficult. The iterative ensemble smoothers utilize ensemble realizations to approximate the first- and second-order moments, which avoids the need to compute derivatives directly. Using the ensemble methods, the form of Eq.~\eqref{eq:rml} in \cite{chen:13} is given by the following update step
\begin{equation}\label{eq:IES}
\begin{aligned}
\bm x_{l+1}&=\bm x_l-\frac{1}{1+\lambda_l}\Delta \bm x_l(\Delta \bm x_l)^T\bm C_x^{-1}
( \bm x_l-\bm x^*) -\Delta \bm x_l(\Delta \bm d_l)^T\Big((1+\lambda_l)\bm C_d+\Delta \bm d_l(\Delta \bm d_l)^T\Big)^{-1} \\
       & \times \Big(g(\bm m_l)-\bm \delta^*-\frac{1}{1+\lambda_l}\Delta \bm d_l(\Delta \bm x_l)^T\bm C_x^{-1}
(\bm x_l-\bm x^*)\Big),
\end{aligned}
\end{equation}
where
\begin{equation*}\left\{
\begin{aligned}
&\Delta \bm x_l=\frac{1}{\sqrt{N_e-1}}(\bm x^1_l,\cdots,\bm x^{N_e}_l)\Big(\bm I_{N_e}-\frac{1}{N_e}\bm 1_{N_e}\bm 1_{N_e}^T\Big),\\
&\Delta \bm d_l=\frac{1}{\sqrt{N_e-1}}\Big(g(\bm m^1_l),\cdots,g(\bm m^{N_e}_l)\Big)\Big(\bm I_{N_e}-\frac{1}{N_e}\bm 1_{N_e}\bm 1_{N_e}^T\Big),\\
&\bm \delta^*=\bm d^o+\bm C_d^{1/2}\bm\epsilon^*, \quad \bm\epsilon^*\sim {\rm N}(\bm 0,\bm I_{N_d}).
\end{aligned}
\right.
\end{equation*}
$N_e$ is the number of samples for the initial ensemble. Here the ensemble samples are used to approximate the gradient of the forward operator $g$ with respect to $x$. The update in Eq.~\eqref{eq:IES} is restricted to space spanned by the initial ensemble and the number of degrees of freedom available for calibration is $N_e-1$. To avoid of the restriction on degrees of freedom, localization is almost always used in high-dimensional problems. For highly nonlinear problems, the ensemble average sensitivity in Eq.~\eqref{eq:IES} is a poor approximation to the local sensitivity, which results in the failure to converge to local minima.


\subsubsection{Hybrid iterative ensemble smoother}
One disadvantage of the RML-IES methods is the use of the same gradient $\bm G$ of the forward operator $g$ with respect to $x$ for each sample. 
To obtain different gain matrices for each sample and avoid the computation of the adjoint systems, we use a hybrid IES method.
For the hybrid IES, instead of using an ensemble stochastic approximation of the sensitivity matrix $\bm G$, we compute the derivative  of $\bm m$ with respective to $\bm x$ analytically and use the chain rule to compute $\bm G$ as 
\[
\bm  G  = \bm G_{m} \cdot   \left( \nabla_{x}  \big(\bm m^T \big)\right)^T = \bm G_m \bm M_x.
\]

Then Eq.~\eqref{eq:IES} can be written as
\begin{equation}\label{eq:hIES}
\begin{aligned}
\bm x_{l+1}&=\bm x_l-\frac{1}{1+\lambda_l}  (\bm x_l-\bm x^*) \\
       &-\bm C_{x}\bm M_{x}^T(\Delta \bm m_l)^{-T}(\Delta \bm d_l)^{T}\Big((1+\lambda_l)\bm C_d+(\Delta \bm d_l)(\Delta \bm m_l)^{-1}\bm M_{x}\bm C_x\bm M_{x}^T(\Delta \bm m_l)^{-T}(\Delta \bm d_l)^{T}\Big)^{-1}\\
       & \times\Bigg(g(\bm m_l)-\bm \delta^*-\frac{1}{1+\lambda_l}(\Delta \bm d_l)(\Delta \bm m_l)^{-1}\bm M_{x} ( \bm x_l-\bm x^* ) \Bigg),
       \end{aligned}
\end{equation}
where $\Delta \bm m_l$ is similar to $\Delta \bm x_l$ and $\bm M_{x}=(\nabla_{x} (\bm m^T))^T$. The gain matrix of each sample is different because of the introduction of the sensitivity matrix $\bm M_x$. Eq.~\eqref{eq:hIES} providing local information of each sample for  the update of parameters. The main disadvantage of a hybrid IES methodology as used previously \cite{oliver:22a} is the large cost of forming and multiplying by the matrix $\bm M_x$  for all samples.


\subsubsection{Nonnegative definite minimal embeddings in circulant matrices}\label{sec:cir_embd}
To accelerate the update step in the  hybrid IES in large inverse problems, we apply circulant embedding for fast multiplication of Toeplitz matrices. Assuming that the forward model is defined on a uniform $(n_x+1)\times (n_y+1)$ grid. The dimension of the corresponding covariance matrix is $(n_x+1)^2\times (n_y+1)^2$, which is extremely large in typical geoscience data assimilation applications. If the prior model covariance function is stationary and the model variables are defined  on a 2D regular, equispaced grid, the covariance matrix $\bm C_x$ is symmetric level-2 block Toeplitz. To reduce the storage and the computation cost, the embedding Toeplitz covariance is applied to circulant matrices. Circulant matrix-vector products can then be computed efficiently by Fast Fourier transform (FFT). A circulant matrix $\check{\bm C}$ is a Toeplitz matrix that has its first column $\check{\bm c}$ periodic. A Toeplitz matrix $\bm C_x$ can always be augmented to generate  a circulant matrix $\check{\bm C}$. This process is called embedding, which can be written as
\[
\bm C_x= {\mathcal M^\dag}\check{\bm C}{\mathcal M}({\mathcal H(\bm\xi)})={\mathcal M^\dag}{\mathcal F}^{[-2]}\Big({\mathcal F}^{[2]}(\check{\bm c}){\mathcal F}^{[2]}(\check{\bm\xi})\Big),
\]
where the operation ${\mathcal M}({\mathcal H(\cdot)})$ injects and embeds $\bm\xi$ into $\check{\bm\xi}$, and $\mathcal F^{[2]}$ is evaluated by FFT. 
Under the assumption of  stationarity, the random field $Y(x,y)$ has correlation function $r(x,y)$ that depends only on the separation of variables. 
Let $h_x$ and $h_y$ be constants denoting, respectively, the horizontal and vertical mesh size of a 2D rectangular domain formed by the points $(x_i,y_j)$ , where
\[\left\{
\begin{aligned}
&x_i=ih_x,\quad 0\leq i\leq n_x\\
&y_j=jh_y,\quad 0\leq j\leq n_y.
\end{aligned}\right.
\]
The ordering of the grid nodes is done from left to right, bottom to top. Thus the correlation matrix $\bm R$ is block Toeplitz \cite{zimmerman:89, dietrich:97}, which is symmetric and uniquely characterized by the first block row
\[
(\bm R_0,\bm R_1,\cdots,\bm R_{n_y})
\]
with $\bm R_j$ being square of dimension $n_x+1$. Based on the node ordering, $\bm R_j$ has first row and first column entries, respectively, given by
\[
{r(ih_x,jh_y)}_{i=0}^{n_x} \quad \text{and} \quad {r(-ih_x,jh_y)}_{i=0}^{n_x}.
\]
The blocks $\bm R_j$ are symmetric only if the correlation function has the special form $r(|x|, |y|)$. When $\bm R_j$ is Toeplitz for $j\geq1$, it is uniquely characterized by its first row and first column which  are written as
\[
(r_{0j},r_{1j},\cdots,r_{n_xj})\quad \text{and} \quad (r_{0j},r_{-1j},\cdots,r_{-n_xj})^T.
\]

The minimal circulant embedding of $\bm R_j$ that ensures that the embedding matrix has an even dimension (for FFT computation) is then given by the square circulant matrix $\bm S_j$ of dimension $2(n_x + 1)$, for which the first row is
\begin{equation}\label{eq:cir_embedding}
(r_{0j},r_{1j},\cdots,r_{n_xj},\phi_j,r_{-n_xj},\cdots,r_{-1j}).
\end{equation}
The circulant embedding of Toeplitz matrices can be constructed by Eq.~\eqref{eq:cir_embedding}. Once constructed, the circulant embedding is used to quickly generate samples from the prior and to evaluate products such as $\bm C_x \bm G^T$, which would otherwise be infeasible.


\subsection{Weighting of ensemble samples}
The RML method of sampling the posterior is only exact if the relationship between the data and the model parameters is linear. For many non-linear problems, however, it is necessary to weight the samples to approximate the posterior distribution, in which case the computation of the gradient of the objective function is necessary. To avoid the need of computing $\bm G$ directly, ensemble-based method  offer an alternative. However, exact sampling using RML requires computation of additional critical points and weighting of solutions \cite{ba:22}. 
The importance weight for the $k$th RML sample is 
\[
w_k\propto \frac{\pi_{\bm X}(\bm x_k)\pi_{\bm\Delta}(\bm\delta_k|\bm x_k)}{p_{\bm X\bm\Delta}(\bm x_k,\bm\delta_k)},
\]
where  $\pi_{\bm X}(\bm x)$ is the prior and the likelihood $\pi_{\bm\Delta}(\bm\delta|\bm x)$ and the proposal density  $p_{\bm X\bm\Delta}(\bm x,\bm\delta)$ is provided by Eq.~\eqref{eq:prop}. Ba et al.\ \cite{ba:22} showed that in high-dimensional nonlinear cases where it is not feasible to sample all critical points, it is valid to randomly sample a single critical point in which case
the factor $n(\bm x')$ in Eq.~\eqref{eq:prop} is set to 1. Finally, the weight of a sample is 
\begin{equation}\label{eq:wei}
w\propto|\bm V|^{1/2} \exp \left[ -\frac{1}{2} \bm \eta(\bm x)\trp \bm V\inv \bm \eta(\bm x) \right] J^{-1}(\bm x,\bm \delta)
\end{equation}
where second derivatives of $\bm G$ at the critical point have been neglected and
\begin{equation*}\label{eq:weight}
\begin{aligned}
J(\bm x,\bm \delta) & \approx|\bm I_{N_x}+\bm C_x\bm G^T\bm C_d\inv \bm G|\\
\bm V(\bm x) & =  \bm C_d+\bm G \bm C_x \bm G\trp \\
 \bm \eta(\bm x) & =g(\bm m) -\bm d^o- \bm G (\bm x-{\bm x}^\text{pr}).
 \end{aligned}
\end{equation*}


\subsubsection{Importance weights for the IES}
Although the IES method is based on RML, the application to multimodal distributions is limited as all samples share a common estimate of $\bm G$ estimated from the ensemble of realizations 
\begin{equation}\label{eq:G_IES}
\begin{aligned}
\bm G\bm C_x\bm G^T & \approx\Delta \bm d\Delta \bm d^T\\
\bm G & \approx\Delta \bm d\Delta \bm x^{T} \bm C_x\inv\\
\bm C_x\bm G^T & \approx\Delta \bm x\Delta \bm d^{T}.
\end{aligned}
\end{equation}
Because $\bm V(\bm x)$ and $J(\bm x,\bm \delta)$ are the same for all samples, the computation of weights can be simplified.  
Neglecting the common multiplying constant,  the IES approximation to the importance weight is 
\begin{equation*}
w \propto \exp \left[ -\frac{1}{2} \bm \eta(\bm x)\trp \bm V\inv \bm \eta(\bm x) \right].
\label{eq:IES_weights}
\end{equation*}
The only difference in weights is a result of differences in the term $\bm\eta(\bm x)$ which requires computation of $\bm G$ from Eq.~\eqref{eq:G_IES}. For most practical problems, the ensemble size is smaller than the dimension of $\bm x$ so the pseudo-inverse is used to approximate the inverse of prior covariance matrix $\bm C_x\inv$.


\subsubsection{Importance weights for the hybrid IES}
The hybrid IES is also based on the RML method of sampling, but uses a different gain matrix for each sample while still avoiding the need for solving the adjoint system. For exactly sampling from the posterior in strongly nonlinear problems, the computation of weights is unavoidable. To compute the weights $\{w_i\}_{i=1}^{N_e}$ of samples generated by hybrid IES, the analytic sensitivity matrix $\bm M_x$ is used, which is $N_m\times N_x$.  The derivative $\bm G_m$ of the objective function with respect to the intermediate variable $\bm m$ can be approximated by the ensemble samples. Finally, the computation of weights in Eq.~\eqref{eq:wei} can be performed by the following forms
\begin{equation*}
\begin{aligned}
\bm G\bm C_x\bm G^T & \approx\Delta \bm d\Delta \bm m\inv\bm M_x\bm C_x\bm M_x^T\Delta \bm m^{-T}\Delta \bm d^T\\
\bm G & \approx \Delta \bm d \Delta \bm m\inv \bm M_x\\
\bm C_x\bm G^T & \approx \bm C_x\bm M_x^T\Delta \bm m^{-T}\Delta \bm d^T.
\end{aligned}
\end{equation*}
The dimension of $\Delta \bm m$ is $N_m\times N_e$. Thus the pseudo-inverse of $\Delta \bm m$ is used in the computation of $\bm G$. For each sample, the terms of Eq.~\eqref{eq:wei} are different. When $\bm M_x$ or $\bm C_x$ has the Toeplitz properties, circulant embedding described  in Sec.~\ref{sec:cir_embd} is used to reduce the computation cost of the matrix multiplication.


\subsection{Excess variance of importance weights}  

For highly nonlinear sampling problems, the variance in the log-weights should be expected to be large since the RML proposal density is not identical to the target density.  On the other hand, the actual spread in computed importance weights is larger than it should be for a number of reasons, including the fact that the minimization method used for computation of samples is approximate. The iterations are generally stopped before actual convergence and the gradient is approximated from a low rank ensemble. A different initial ensemble would result in a different final estimate of $\bm G$, $\bm V$ and $\det {\bm V}$. 
All of these will result in variability in computation of weights and the un-normalized log-weights will consequently have a large spread. If the ``noisy'' log-weights are used directly to compute weighted forecasts, almost all the weight will fall on a single realization.

For large Bayesian inverse problems of the type encountered in the geosciences, the likelihood is often difficult to evaluate and noisy approximations to the likelihood must instead be used \cite{dunbar:22}. When the likelihood is noisy, however, the transition kernel in MCMC or equivalently the weighting of particles in important sampling will be affected by the noise \cite{alquier:16,acerbi:20}. This noise must either be removed or be otherwise accounted for if the sampling is to be efficient.
The problem of sampling with noisy importance weights has  been reviewed by \cite{akyildiz:17}, who showed that denoising can be an effective approach.
In the application to weighting of RML samples, the errors appear primarily in the evaluation of the proposal density, not in the evaluation of the likelihood as in most previous studies. Because the importance weights are ratios of likelihood to proposal density, however, the effect of noise in either term on the weight is similar.

\subsubsection{A model for noise in the log-weights}\label{subsec:denosing}  

For simplicity, we denote the logarithm of the weights on the particles as $\omega$ so that
\begin{equation*}
\omega  = -\frac{1}{2} \log \det \bm V - \frac{1}{2}  \bm\eta\trp \bm V\inv \bm\eta
\end{equation*}
where 
\begin{equation*}
\bm V = \bm C_d + \bm G \bm C_x \bm G\trp
\end{equation*}
and 
\begin{equation*}
\bm\eta (\bm x)  =  g(\bm x) - \bm d^o - \bm G (\bm x- \bm x\spr{pr}).
\end{equation*}

For a non-linear problem the sensitivities $\bm G$ at the minimizer will be variable, and since $\bm G$ and the data mismatch enter quadratically in $\omega$, we expect the ``true'' distribution of log-weights to be approximately  chi-square. 
We additionally assumed a Gaussian model for the distribution of errors in the computed log-weights, which we also refer to as noise in the weights.

\begin{figure}[htbp]
\begin{subfigure}{0.31\textwidth}
\includegraphics[width=0.95\textwidth]{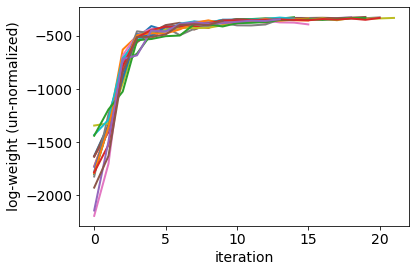}
\caption{Evolution of log-weights of a single particle in 16 independent ensembles.}
\label{fig:obs_and_prior_distributions_mono_a}
\end{subfigure}
\hfill
\begin{subfigure}{0.31\textwidth}
\includegraphics[width=0.95\textwidth]{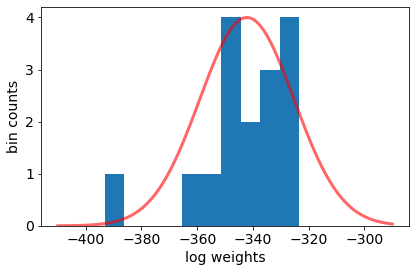} 
\caption{Distribution of final log-weights for a unique sample from the prior and Gaussian fit.}
\label{fig:obs_and_prior_distributions_mono_b}
\end{subfigure} 
\hfill
\begin{subfigure}{0.31\textwidth}
\includegraphics[width=0.95\textwidth]{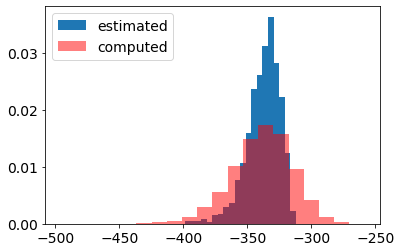} 
\caption{Computed distribution of all log-weights from 8 ensembles of 200 realizations.}
\label{fig:obs_and_prior_distributions_mono_c}
\end{subfigure} 
\caption{The distributions of log-weights for the monotonic log-permeability transform. }
\label{fig:obs_and_prior_distributions_mono}
\end{figure}

We can obtain an empirical characterization of the computation error  by generating a number of realizations of the computed value of $\omega$ for the same prior sample $\bm x^\ast$ but with different ensembles of realizations used for computation of $\bm G$. We did  this for realizations of the monotonic transform of log-permeability by generating  16 independent ensembles of 199 realizations and augmenting each ensemble with another realization that was then common to each ensemble. Figure~\ref{fig:obs_and_prior_distributions_mono_a} shows the evolution of the log-weight on the single realization that was common to all 16 ensembles. The minimization was stopped when the iterations reach the terminated condition. (The particle that was common to all ensemble members stopped updating by iteration 10 in all ensembles, where the iteration number just contains the iteration that the mean of data mismatch is smaller than last iteration.) Figure~\ref{fig:obs_and_prior_distributions_mono_b} shows the distribution of final values of $\omega$ (blue) and the Gaussian fit to the distribution of final values (red curve).

Estimating reasonable parameters in the chi-square model of the true distribution for large weights is more difficult than estimating the errors in the computation, partly because we do not have an empirical distribution of log-weights without computational noise.  We instead  used a trial-and-error approach in which the observed distribution of log-weights was compared with a Monte Carlo distribution of noisy samples from a chi-square distribution whose  parameters were tuned to match the observed distribution. 
Figure~\ref{fig:obs_and_prior_distributions_mono_c}  compares the distribution of RML computed realisations with the realisations from the modelled distribution of noisy large weights.

Once the parameters of the error model and the parameters of the true log-weights have been estimated, it is straightforward to denoise to computed values of $\omega$. In this study, for each observed  value $\omega^o$ we compute the maximum a posteriori estimate of $\omega$.
\begin{equation*} 
P(\omega | \omega^o)  \propto  P(\omega^o | \omega) P(\omega) 
\end{equation*} 
or
\begin{equation*}
P(\omega | \omega^o)  \propto  
\begin{cases} \exp \left(- \frac{(\omega-\omega^{o})^2}{2\sigma_o^2}\right) \left( \frac{\omega-\omega^{pr}}{\sigma_{pr}} \right)^{\nu/2-1} \exp \left(- \frac{\omega-\omega^{pr}}{2  \sigma_{pr} } \right)  \quad &  {\rm for}  \quad \omega > \omega_{pr} \\
0 \quad & {\rm for}  \quad \omega \le \omega_{pr}.  \end{cases}
\end{equation*}

\begin{figure}[htbp]
\begin{subfigure}{0.45\textwidth}
\includegraphics[width=0.95\textwidth]{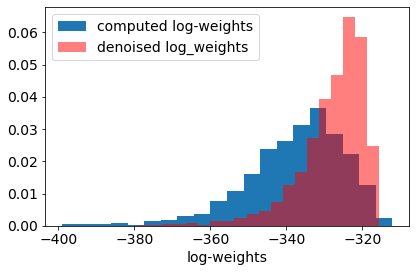}
\caption{Distribution of log-weights after denoising for monotonic log-permeability  transformation.}
\label{fig:denosied_distributions_a}
\end{subfigure} 
\hfill
\begin{subfigure}{0.45\textwidth}
\includegraphics[width=0.95\textwidth]{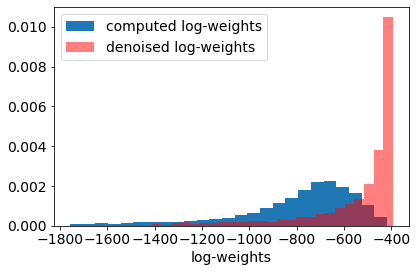}
\caption{Distribution of log-weights after denoising for non-monotonic log-permeability transformation.}
\label{fig:denosied_distributions_b}
\end{subfigure} 
\caption{The distributions of log-weights after denoising (red colors). }
\label{fig:denosied_distributions}
\end{figure}

For the monotonic log-permeability transform, with  $\sigma_o = 16.9$, $\sigma_{pr} = 6$, and $\nu = 4$ we obtain the denoised log-weights shown in red in Fig.~\ref{fig:denosied_distributions_a}. The effective sampling efficiency, $N\sbr{eff}/N_e = 97.8/1600 \approx 0.109$, based on Kong's estimator Eq.~\eqref{eq:NEff},
\begin{equation}\label{eq:NEff}
N_{\text{Eff}}=\frac{1}{\sum_{k=1}^{N_e}w_k^2},
\end{equation}
where $\sum_{k=1}^{N_e}w_k=1$.

The spread in the weights for the case with non-monotonic transform of log-permeability is much larger than in the case with monotonic permeability transform. First,  the computation of weights appears to less repeatable: Fig.~\ref{fig:denosied_distributions_a}  shows the evolution of un-normalized log-weights for the same sample when included in 16 otherwise independent  ensembles. The spread of the final values for the common particle (Fig.~\ref{fig:denosied_distributions_b}) is approximately 5 times larger for the non-monotonic case ( $\sigma_o = 95.3$) than for the monotonic case ($\sigma_o = 16.9$). Presumably, the additional variability is a result of greater variability in $\bm G$ and the presence of more local minima. Additionally, the prior spread of log-weights appears to be larger, again because of multiple minima and the fact that the proposal distribution is farther from the target distribution in this case. For the non-monotonic log-permeability transform, with  $\sigma_o = 95.3$, $\sigma_{pr} = 13$, and $\nu = 3$ we obtain the denoised log-weights shown in red in Fig.~\ref{fig:denosied_distributions_b}. The effective sampling efficiency, $N\sbr{eff}/N_e = 27.2/1600 \approx 0.017$, based on Kong's estimator Eq.~\eqref{eq:NEff}.

\begin{figure}[htbp]
\begin{subfigure}{0.31\textwidth}
\includegraphics[width=0.95\textwidth]{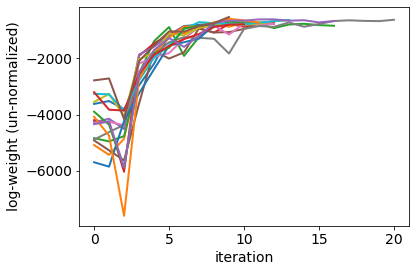}
\caption{Evolution of log-weights of a single particle in 16 independent ensembles.}
\label{fig:obs_and_prior_distributions_nonmono_a}
\end{subfigure}
\hfill
\begin{subfigure}{0.31\textwidth}
\includegraphics[width=0.95\textwidth]{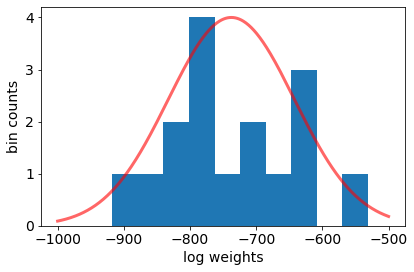} 
\caption{Distribution of final log-weights for a unique sample from the prior and Gaussian fit.}
\label{fig:obs_and_prior_distributions_nonmono_b}
\end{subfigure} 
\hfill
\begin{subfigure}{0.31\textwidth}
\includegraphics[width=0.95\textwidth]{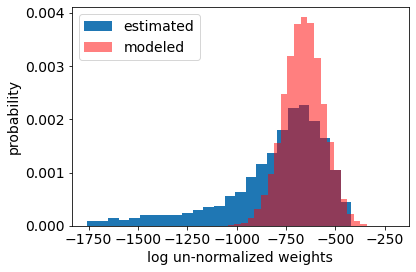} 
\caption{Computed distribution of all log-weights from 8 ensembles of 200 realizations.}
\label{fig:obs_and_prior_distributions_nonmono_c}
\end{subfigure} 
\caption{The distributions of log-weights for the non-monotonic log-permeability transform. }
\label{fig:obs_and_prior_distributions_nonmono}
\end{figure}

\section{Weighted RML applications}

In this section, two data assimilation methods (hybrid IES and IES) are applied to a 2-dimensional, 2 fluid phase flow problem with  permeability transforms, $\bm m=f(\bm x)$, of varying degrees of non-linearity. 
We first demonstrate the hybrid IES methodology for the monotonic and non-monotonic log-permeability transforms. The sensitivity of the log-permeability field  ($\bm m$) to the correlated multi-normal variables $\bm x$ are computed analytically. The derivatives of production data with respect to the log-permeability variables are estimated from the ensemble of realizations. We then present results for the application of the IES method to the same problems, in this case, however, the derivatives of production data to the $\bm x$ are estimated directly from the ensemble. For the two applications, the uncertain permeability field in a 2D porous medium is estimated by assimilation from a time series of water cut observations at 9 producing wells. The state, $u$, of an incompressible and immiscible two-phase flow system is determined by the pressure $p(x,t)$ and  saturation $s(x,t)$, which in this example are governed by 
\begin{equation}\label{eq:two_phase}
\left\{
\begin{aligned}
-\nabla\cdot({\mathbf K}\lambda_{*}(s)\nabla p)&=q,\\
\phi\frac{\partial s}{\partial t}+\nabla\cdot(f_{*}(s) v)&=\frac{q_w}{\rho_w}\quad \text{in} \quad \Omega\times [0,T]
\end{aligned}\right.
\end{equation}
with the mixed boundary condition
\[
v_w\cdot \mathbf n=0\quad \text{on} \quad \partial\Omega \times [0,T], \quad s(x,0) = 0 \quad\text{in}\quad \Omega=[0,2]\times[0,2],
\]
where
\begin{equation}
\begin{aligned}
\frac{q_w}{\rho_w}  & = \max(q,0) +f_{*}(s)\min(q,0),\\
 v & = -\vc K\lambda\nabla p \\ 
 q & = \frac{q_w}{\rho_w}+\frac{q_o}{\rho_o},\quad p_w=p_o,\quad \lambda_{*}(s) = \frac{k_{ri}}{\mu_i},\quad i=w,o
\end{aligned}
\label{eq:porous_flow}
\end{equation}
and $k_{ri}$ and $\mu_i$ are the relative permeability and viscosity of phase $i$ respectively.

\begin{figure}[htbp]    
\begin{tabular}{ccc}
\includegraphics[width=0.31\textwidth]{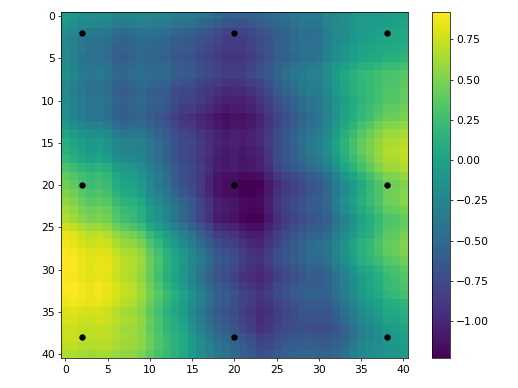}
 &
 \includegraphics[width=0.31\textwidth]{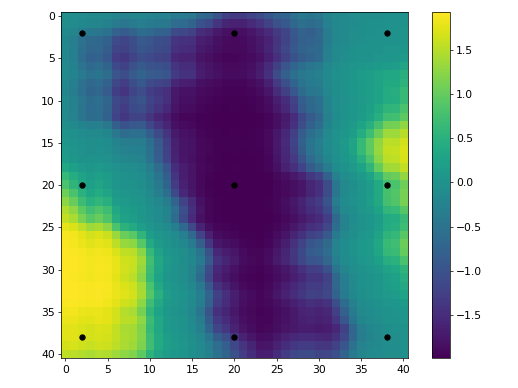}
 &
\includegraphics[width=0.31\textwidth]{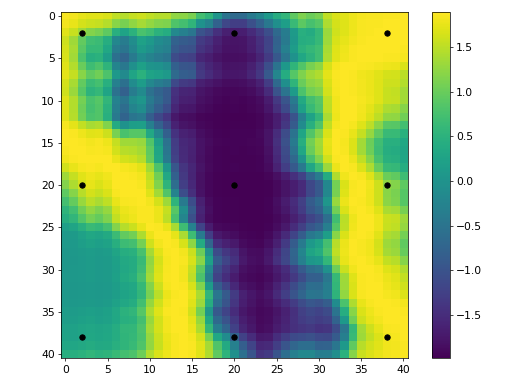} 
 \\
(a) identity transform & (b) monotonic transform & (c) non-monotonic transform
\end{tabular}
\caption{The ``true'' log-permeability fields used to generate production data and forecasts. Black dots show locations of producing wells.}
\label{fig:true_logK}
\end{figure}

\begin{figure}[htbp]     
\begin{tabular}{ccc}
\includegraphics[width=0.45\textwidth]{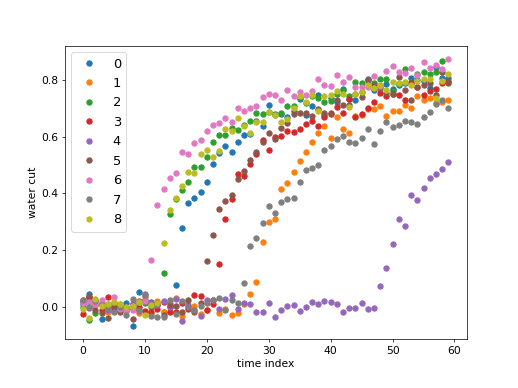}
 &
 \includegraphics[width=0.45\textwidth]{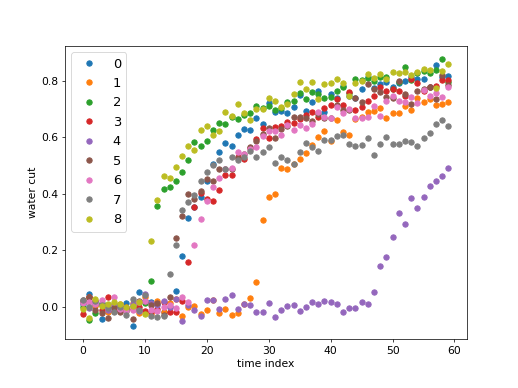}
 \\
(a) monotonic transform & (b) non-monotonic transform
\end{tabular}
\caption{Noisy observations of water cut in 9 producing wells.}
\label{fig:Obser}
\end{figure}

It is not possible to model permeability  as a Gaussian random variable because it is required to be non-negative. We evaluate sampling with two different prior distributions for permeability. In both cases, we define a latent variable $\bm x$ that is multivariate Gaussian with covariance
\begin{equation}\label{non-permeability}
C(x,y)=\sigma^2 \left(1-\frac{x^2+y^2}{\rho^2}\right)\exp \left(-\frac{x^2+y^2}{\rho^2}\right).
\end{equation}
The permeability field for the data-generating model  is a draw from a prior model with the range parameter for the correlation length $\rho=1.1$ and  standard deviation 0.8 for the monotonic and non-monotonic transforms. The true data-generating permeabilities are as shown in Fig.~\ref{fig:true_logK}. Figure~\ref{fig:Obser} displays the water cut observations from the 9 producing wells for both permeability transforms. To compare the results from standard IES and hybrid IES, ensemble size $N_e$ of the two methods is set as 200. 

The observation locations are distributed on uniform $3 \times 3$ grid of the domain $[0.1, 1.9] \times [0.1, 1.9]$ as shown in Fig.~\ref{fig:true_logK} (black dots). The noise in the observations is assumed to be Gaussian and independent with standard deviation 0.02. The forward model (Eq.~\eqref{eq:porous_flow}) is solved by the two-point flux-approximation (TPFA) scheme, which is a cell-centred finite-volume method \cite{aarnes:07}. For the two test cases, the forward model is defined on a uniform $41 \times 41$ grid with time step $\Delta t=0.1$. The dimension of the discrete parameter space is 1681.


\subsection{History matching using hybrid IES for problems with non-linear permeability transformation}\label{sec:hIES_for_hm}
For the hybrid IES method, the gradient $\bm M_x$  of log-permeability $m$ with respect to the Gaussian parameter $x$ is necessary. We selected two permeability transforms that have characteristics similar to rock ``facies'' distributions, i.e.\ regions with relatively uniform permeability and sharp transitions to a different facies.

\begin{figure}[htbp]    
\begin{tabular}{p{0.2em}c|c}
\raisebox{1.5ex}{\rotatebox{90}{prior}} &
\includegraphics[width=0.45\textwidth]{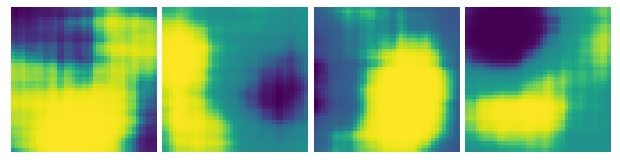} &
\includegraphics[width=0.45\textwidth]{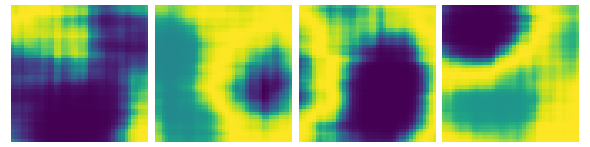}
\\
\raisebox{0.5ex}{\rotatebox{90}{posterior}} &
\includegraphics[width=0.45\textwidth]{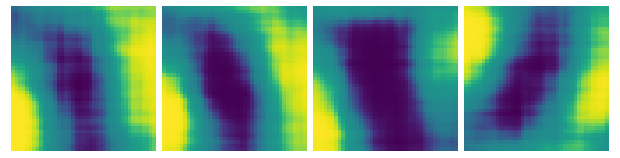} &
\includegraphics[width=0.45\textwidth]{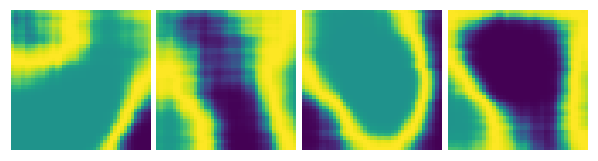}
\\ \hline
\raisebox{0.5ex}{\rotatebox{90}{weights}} &
\begin{overpic}[width=0.45\textwidth]{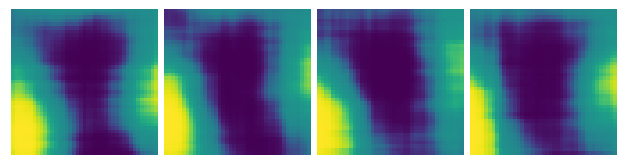} 
\put(-12,2.5){\rotatebox{90}{largest}}
\end{overpic} &
\includegraphics[width=0.45\textwidth]{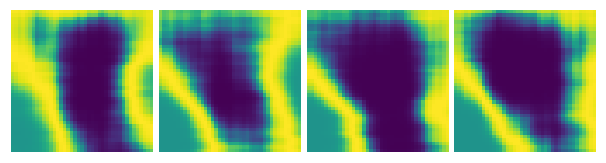} 
\\
   &   (a) monotonic   &  (b) non-monotonic 
\end{tabular}
\caption{Model realizations for monotonic and non-monotonic transformations of log-permeability using hybrid IES.}
\label{fig:monotonic_non-monotonic_hIES}
\end{figure}

\subsubsection{Monotonic log-permeability transform}
In some cases, the region occupied by the high permeability facies is typically isolated and  can be modelled well by a monotonic transformation of a Gaussian variable to log-permeability. To illustrate the effect of this type of nonlinearity,  the following transformation is applied
\begin{equation}\label{eq:monotonic}
m =  \tanh \big(4 x +2 \big)  +   \tanh \big(4x-2\big),
\end{equation}
where $x$ is the prior Gaussian random variable.  The analytic derivative is given by
\begin{equation*}
\frac{dm}{dx} = 8 - 4  \tanh^2 \big(4 x +2 \big) -   4\tanh^2 \big(4x-2\big).
\end{equation*}
In this case,  the spatial distribution of rock types is obtained by applying a soft threshold to a latent Gaussian random field. With this transformation, values of $x <1$ are assigned $m \approx -2$ and values values of $x > 1$ are assigned $m \approx 2$.
The discretized form of the sensitivity ${\bm M}_x$ is diagonal with 
\[ {\bm M}_x = \begin{bmatrix}
      \frac{dm_1}{dx_1} & 0 & \hdots & 0 \\
      0 & \frac{dm_2}{dx_2} &    & 0 \\
      \vdots &   & \ddots & \vdots \\
      0 & 0 & \hdots &  \frac{dm_{N_x}}{dx_{N_x}}
\end{bmatrix}. \]
while the covariance operator $\bm C_x$ of Eq.~\eqref{non-permeability} is dense but block-Toeplitz  \cite{zimmerman:89, dietrich:97}.
Multiplication by ${\bm M}_x$ is trivial, but the product $\bm C_x \bigl({\bm M}_x^T (\Delta \bm m_i)^{-T} \bigr)$ is computed using the Toeplitz property of $\bm C_x$ as described in Sec.~\ref{sec:cir_embd}.

\begin{figure}[htbp]   
\includegraphics[width=0.9\textwidth]{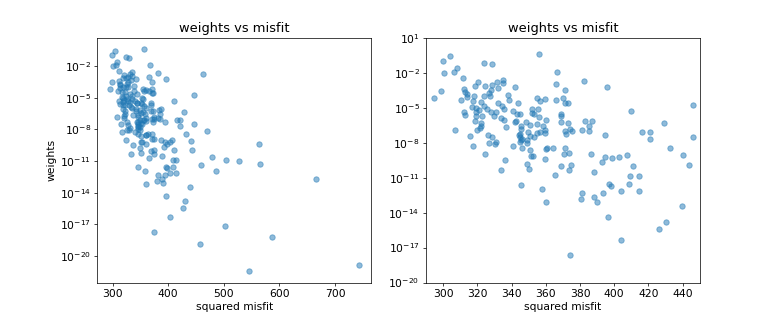} 
\caption{The weights vs misfits for the monotonic log-permeability transform. Blue points show computed weights.}
\label{fig:wei_vs_mis_hIES_mon}
\end{figure}

As the gain matrices are potentially different for each realization in the hybrid IES, we might expect that some realizations will converge to local minima with small probability mass if the posterior has multiple modes.    The samples with largest weights  are  likely to be similar, however.  Figure~\ref{fig:monotonic_non-monotonic_hIES}a shows the log-permeability fields for the first 4 prior realizations (top row) and corresponding posterior realizations of log-permeabilities (middle row). The variability in the posterior realizations is smaller than the variability in the prior realizations, but still fairly large.  In contrast, the log-permeabilities of posterior realizations  with the 4 largest weights (Fig.~\ref{fig:monotonic_non-monotonic_hIES}a (bottom row))  are much more similar. Qualitatively, it appears that   importance weighting is beneficial in correcting the posterior samples.   For nonlinear problems such as this, we also expect the approximate posterior realizations with largest weights to have small data mismatch with observations. To investigate this hypothesis, a crossplot of the weights vs squared data misfits is shown in Fig.~\ref{fig:wei_vs_mis_hIES_mon}.  Although the log-weights are correlated to squared data misfit and the models with largest data misfit have very small weights, there is large variability in weights even for small data misfit.  The more important observation is that the nonlinearity in $g(\cdot)$ increases the variability in weights and the mean of the squared misfit (369) is substantially larger than the expected value for samples from the posterior (270), while the weighted mean is 330.

\begin{figure}[htbp]     
\includegraphics[width=0.9\textwidth]{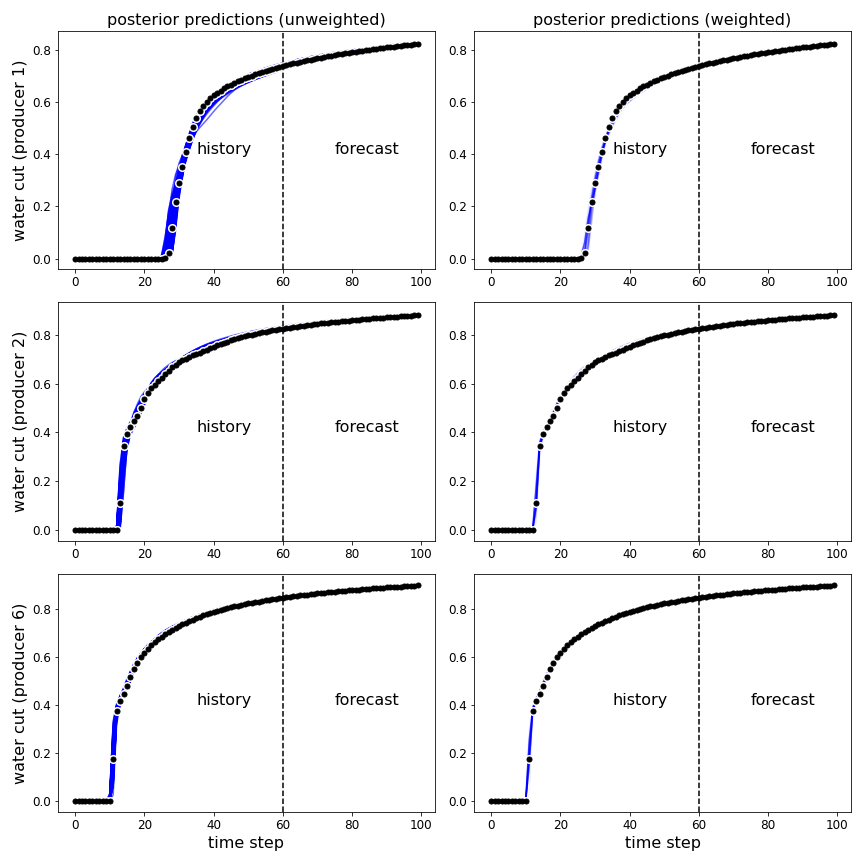} 
\caption{The posterior predictions of producing wells 1, 2, 6 using unweighted and weighted hybrid IES for the monotonic log-permeability transform. Black points show true observations.}
\label{fig:post_predict_hIES_mon}
\end{figure}

Because the range of the covariance of the permeability field is relatively large compared to the domain of interest, the observation locations are spatially distributed, and the production data from all wells are matched fairly well by the weighted and unweighted samples (Fig.~\ref{fig:post_predict_hIES_mon}), the posteriori means of the log-permeability fields  (not shown) look similar  to the truth, except that the truth is somewhat ``rougher''.

The justification for data assimilation or history matching of subsurface models is generally to provide accurate assessments of future reservoir behavior.  Figure~\ref{fig:post_predict_hIES_mon} show the quality of the match to observed data and the predictability of future performance of the unweighted and weighted posterior ensembles at three representative wells. For this case, the differences in predictability between the weighted and unweighted realizations are small, although the prediction interval is narrower for the weighted hybrid IES. This is a result of the large variance of weights, which results in a small effective sample size.

\subsubsection{Non-monotonic log-permeability transform}

To obtain  permeability fields with a low permeability `background´ and connected high perm ‘channels’, we used the  transformation  
\begin{equation}\label{eq:non-monotonic}
m =  2 \tanh \big(4 x +2 \big)  +   \tanh \big(2-4x\big)-1,
\end{equation}
where $x$ is again the prior Gaussian random variable. The corresponding derivative of log-permeability with respect to the Gaussian latent variable is then
\begin{equation*}
\frac{dm}{dx} = 4 - 8  \tanh^2 \big(4 x +2 \big)  +   4\tanh^2 \big(2-4x\big).
\end{equation*} 
The ‘true’ data-generating log-permeability field for this test problem is shown in Fig.~\ref{fig:true_logK}c and water cut observations  for the 9 producing wells are plotted in Fig.~\ref{fig:Obser}b. Although the data are not noticeably different from the data is the monotonic case (Fig.~\ref{fig:Obser}a), the presence of the channel facies makes the problem slower to converge to a mode and more likely to converge to a mode with small probability mass. 

\begin{figure}[htbp]  
\includegraphics[width=0.9\textwidth]{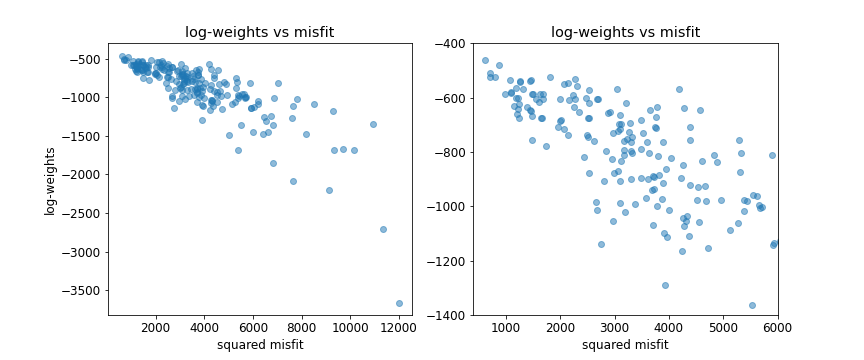} 
\caption{The log-weights vs misfits for the non-monotonic log-permeability transform using hybrid IES. Blue points show computed log-weights.}
\label{fig:wei_vs_mis_hIES_non}
\end{figure}

For the non-monotonic transform case, the difference between posterior realizations is larger than in the monotonic transform case as illustrated by the first 4 prior realizations  and  corresponding posterior realizations (Fig.~\ref{fig:monotonic_non-monotonic_hIES}b). 
In this case, the differences are  due to the use of the analytic sensitivity $\bm M_x$, which allows realizations to converge to different local minima. 
The posterior realizations with the largest importance weights (Fig.~\ref{fig:monotonic_non-monotonic_hIES}b (bottom row)) show reasonable similarity to the true field. 
In addition to the greater diversity in the realizations compared to the monotonic case, the importance weights and the data mismatch are also much more diverse for the non-monotonic transform (Fig.~\ref{fig:wei_vs_mis_hIES_non}.).
In the non-monotonic case, the expected mean of the data misfit part of the log-likelihood is still 270 (half the number of observations). 
Fig.~\ref{fig:wei_vs_mis_hIES_non} shows, however, that the data misfits of most posterior samples are concentrated in the interval $[1000, 6000]$. 
The posterior mean of unweighted data misfits is 3794 -- approximately 14 times larger than the expected value. The posterior mean of data misfits for  weighted  realizations, on the other hand,  is 617, which is still larger than expected, but much smaller than the mean for the unweighted realizations.

\begin{figure}[htbp]    
\begin{tabular}{c|cc|cc}
\includegraphics[width=0.18\textwidth]{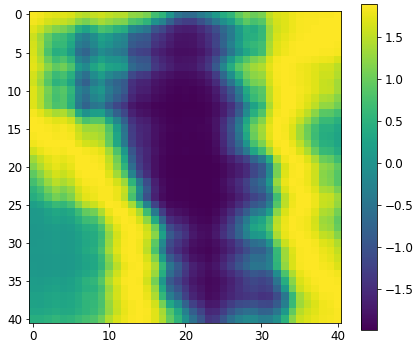}. 
&
\includegraphics[width=0.18\textwidth]{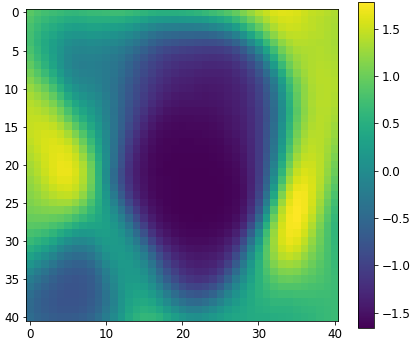}    
&
\includegraphics[width=0.18\textwidth]{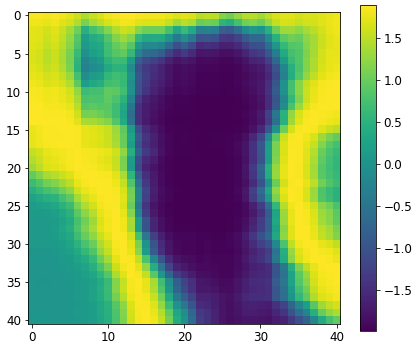}
&
\includegraphics[width=0.18\textwidth]{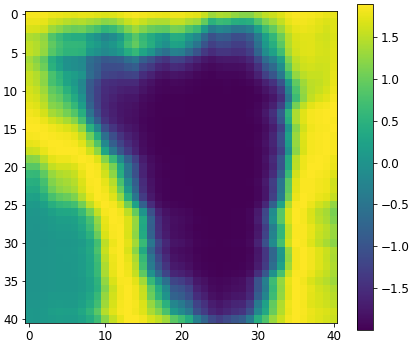} 
&
\includegraphics[width=0.18\textwidth]{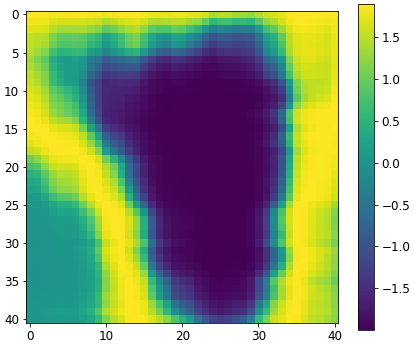}
\\
\includegraphics[width=0.18\textwidth]{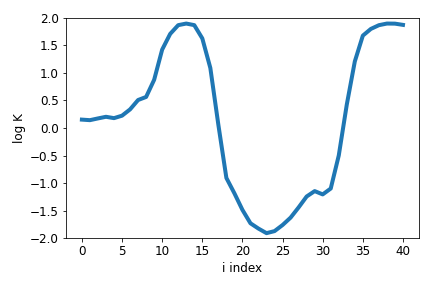}    
&
\includegraphics[width=0.18\textwidth]{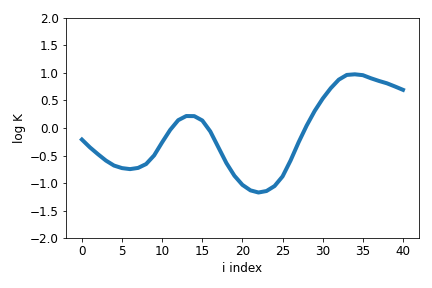}    
&
\includegraphics[width=0.18\textwidth]{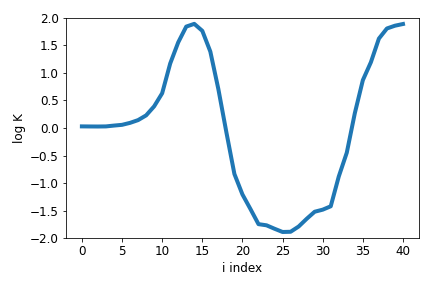}   
&
\includegraphics[width=0.18\textwidth]{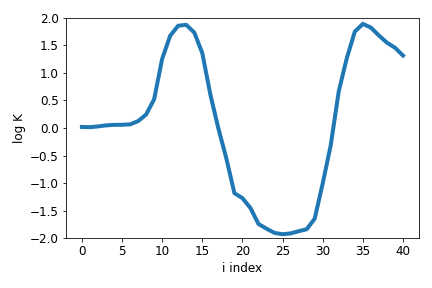}   
&
\includegraphics[width=0.18\textwidth]{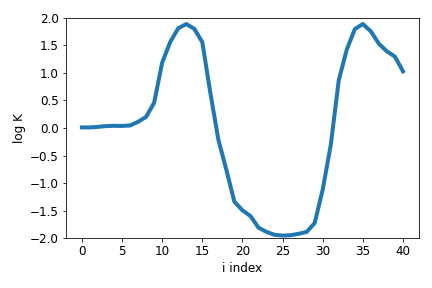}   
\\
\includegraphics[width=0.18\textwidth]{true_log-per_hIES_non_i35.png}    
&
\includegraphics[width=0.18\textwidth]{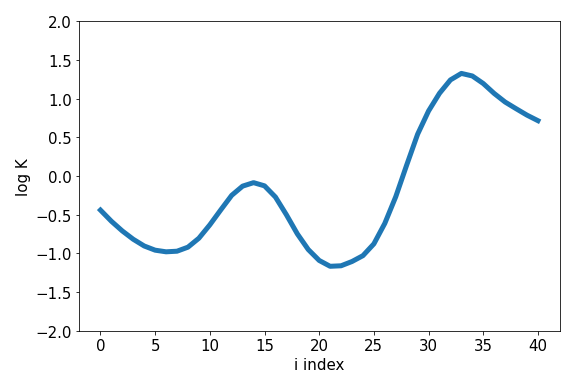}    
&
\includegraphics[width=0.18\textwidth]{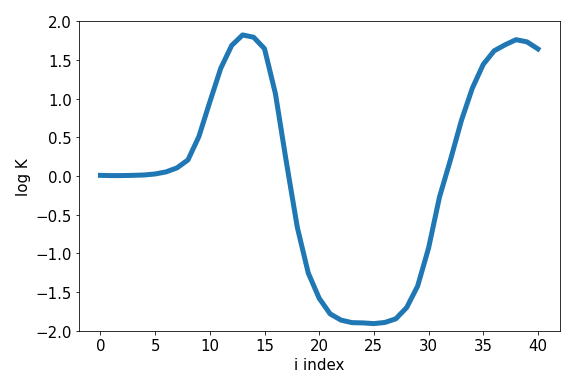}   
&
\includegraphics[width=0.18\textwidth]{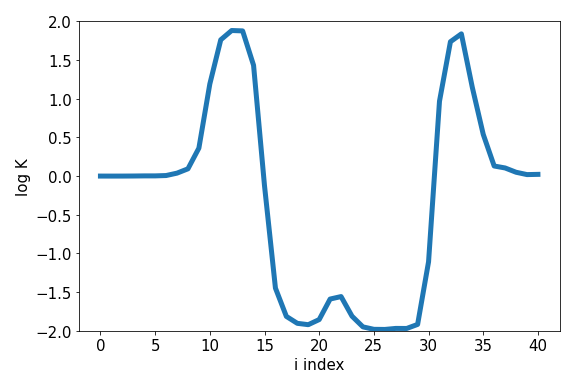}
&
\includegraphics[width=0.18\textwidth]{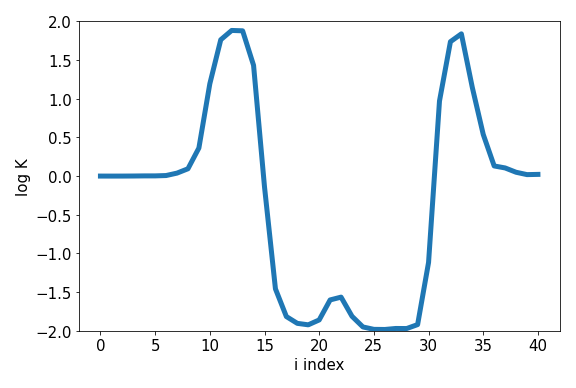}
\\
Truth  &  unweighted  &  weighted   &  unweighted  &  weighted  \\
       &    \multicolumn{2}{c}{hybrid IES} & \multicolumn{2}{|c}{IES}
\end{tabular}
\caption{The true log-permeability (left), unweighted and weighted posterior mean using hybrid IES (middle columns) and standard IES (right) for the non-monotonic transform. In each case, the top row shows the entire mean log-permeability fields, while the middle and  bottom rows shows the  posterior mean values of log-permeability at location $(x,1.75)$ for two different selection of convergence parameters. }
\label{fig:per_post_non_hIES}                
\end{figure}

\begin{figure}[htbp]
\includegraphics[width=0.9\textwidth]{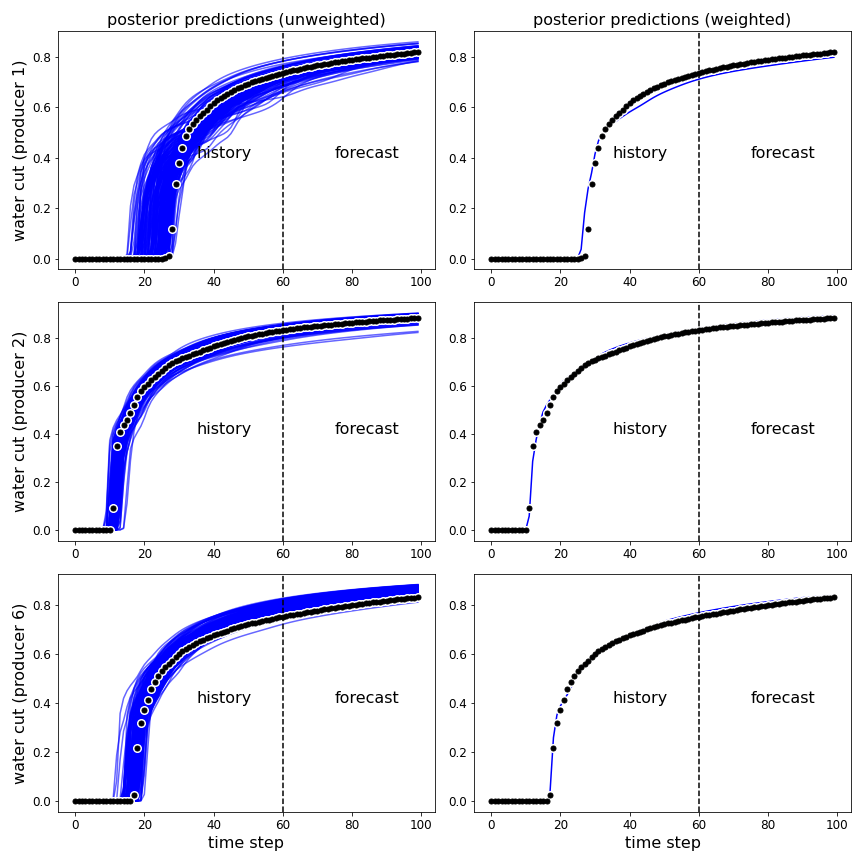} 
\caption{The posterior predictions for producing wells 1, 2, 6 using unweighted and weighted hybrid IES for the non-monotonic log-permeability  transform. Black points show true observations.}
\label{fig:post_predict_hIES_non}
\end{figure}

Weighted and unweighted mean log-permeability fields for hybrid IES assimilation are shown in  Fig.~\ref{fig:per_post_non_hIES}.  The middle and bottom rows of Fig.~\ref{fig:per_post_non_hIES} show mean values for two different sets of results using different convergence trajectories (i.e.\ different values for the rmultiplier of $\lambda$ in Levenberg-Marquardt minimization).
The unweighted means for hybrid IES bear limited similarity to the true field and shows little connectivity of the high permeability facies. This is a result of averaging with many dissimilar realizations that are not all well calibrated. The weighted mean looks much more like the truth as it puts more weight on samples with higher probability mass. The effect of importance weighting is perhaps more obvious in the posterior distribution of predictions of water cut.
The spread in the unweighted predictions is large, even during the history-matched period (Fig.~\ref{fig:post_predict_hIES_non} (left column)) and much larger than expected given the observation error. 
On the other hand, the quality of the  weighted posterior realizations (right column) is excellent, except for well 1. The main problem with the weighted ensemble appears to be too small spread resulting from the small effective sample size.

\subsection{History matching using IES}

In this section, we apply the standard IES methods to the two-phase flow model of Eq.~\eqref{eq:two_phase} and  unweighted and weighted results are compared with results of the hybrid IES of Sec.~\ref{sec:hIES_for_hm}. Data assimilation is performed on  log-permeability fields generated using the monotonic and non-monotonic transforms of Eq.~\eqref{eq:monotonic} and Eq.~\eqref{eq:non-monotonic}. The IES method is computationally simpler than the hybrid IES as  the computation of sensitivity $\bm M_x$ is avoided. As a consequence, however, the standard IES is incapable of sampling multiple minima.  

\begin{figure}[htbp]
\begin{tabular}{p{0.2em}c|c}
\raisebox{1.5ex}{\rotatebox{90}{prior}} &
\includegraphics[width=0.45\textwidth]{prior4_monotonic_2022-11-27_Newseed.png} &
\includegraphics[width=0.45\textwidth]{prior4_non-monotonic_2022-11-27_Newseed.png}
\\
\raisebox{0.ex}{\rotatebox{90}{posterior}} &
\includegraphics[width=0.45\textwidth]{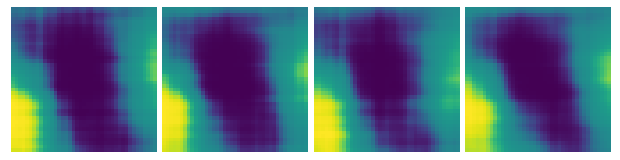} &
\includegraphics[width=0.45\textwidth]{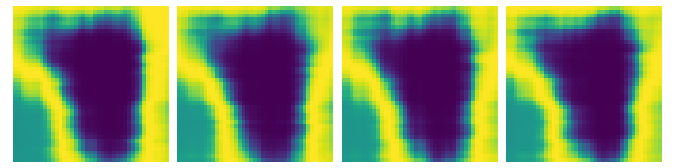}
\\ \hline
\raisebox{0.5ex}{\rotatebox{90}{weights}} &
\begin{overpic}[width=0.45\textwidth]{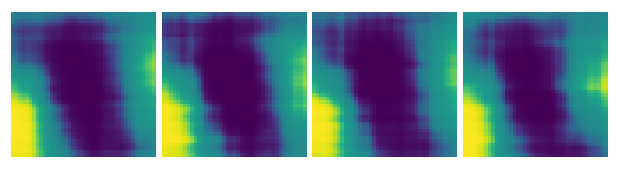} 
\put(-12,2.5){\rotatebox{90}{largest}}
\end{overpic} &
\includegraphics[width=0.45\textwidth]{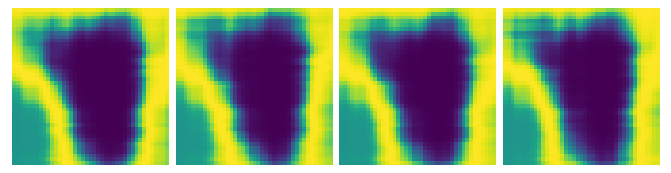} 
\\
   &   (a) monotonic   &  (b) non-monotonic 
\end{tabular}
\caption{Model realizations for monotonic and non-monotonic  transformations of log-permeability using IES.}
\label{fig:monotonic_non-monotonic_IES}
\end{figure}

\begin{figure}[htbp]
\begin{tabular}{ccc}
\raisebox{7ex}{\rotatebox{90}{IES}} 
\includegraphics[width=0.9\textwidth]{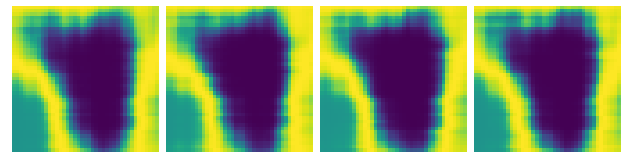}  
\\
\raisebox{5ex}{\rotatebox{90}{hybrid IES}} 
\includegraphics[width=0.9\textwidth]{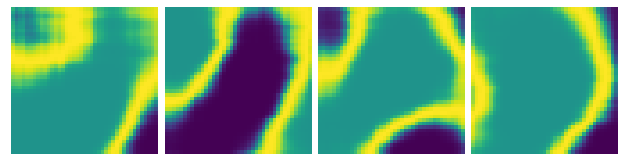} 
\end{tabular}
\caption{Model realizations with the smallest weights for non-monotonic  transformations of log-permeability using IES and hybrid IES.}
\label{fig:poor_post_IES_hIES}
\end{figure}

Because the same prior ensemble of realizations is used for the numerical experiments in Sec.~\ref{sec:hIES_for_hm}  and in this section, the prior log-permeability fields shown in the top row of  Fig.~\ref{fig:monotonic_non-monotonic_hIES}  are the same as shown in the top row of Fig.~\ref{fig:monotonic_non-monotonic_IES}. 
Figure~\ref{fig:monotonic_non-monotonic_IES} shows log-permeability fields for the first 4 posterior realizations (middle row) and the posterior realizations with the largest weights (bottom row) for the two cases (monotonic and non-monotonic) using the IES for data assimilation. 
As the same gain matrix is used for all samples generated from standard IES, the variability among posterior approximate realizations is smaller for the IES than for the hybrid IES. 
Unlike the situation with the hybrid IES, the first four posterior realizations (Fig.~\ref{fig:monotonic_non-monotonic_IES} (middle row)) are similar to the four realizations with the largest weights  (Fig.~\ref{fig:monotonic_non-monotonic_IES} (bottom row)) and  the unweighted and weighted posterior means obtained using standard IES are almost the same. In fact, the IES realizations with the smallest weightes are nearly identical to the realizations with the largest weights (Fig.~\ref{fig:poor_post_IES_hIES}), so importance weighting has very little effect for the IES method on this problem. The effective sample size of the IES for the two cases is low because the posterior spread has been underestimated \cite{chen:17, ba:21}.

\begin{figure}[htbp]
\begin{subfigure}{0.5\textwidth}
\includegraphics[width=0.95\textwidth]{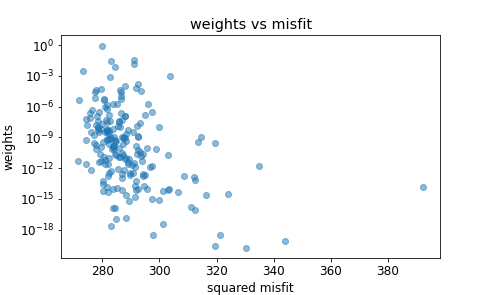}
\caption{monotonic}
\label{fig:wei_vs_mis_IES_a}
\end{subfigure}
\hfill
\begin{subfigure}{0.5\textwidth}
\includegraphics[width=0.95\textwidth]{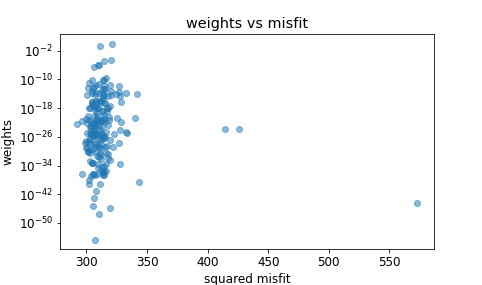} 
\caption{non-monotonic}
\label{fig:wei_vs_mis_IES_b}
\end{subfigure} 
\caption{The weights vs misfit for monotonic and non-monotonic transformations of log-permeability using IES. Blue points show computed weights.}
\label{fig:wei_vs_mis_IES}
\end{figure}

We expect the spread of weights in the IES method to be smaller than the spread of weights in application of the hybrid IES since the IES generated less variety in the posterior realizations.  But in fact the spread of weights is quite large and almost independent of the data mismatch (Fig.~\ref{fig:wei_vs_mis_IES}). This appears to be a result of errors  in computation of $\bm G$ and an underestimate of the magnitude of $\bm V$.

For a Gauss-linear inverse problem there should be no correlation between the weight on a sample from the posterior and the data mismatch -- in fact, for this case, the weights should be uniform when a minimization-based sampling approach is used. For the nonlinear 2D porous flow example with monotonic log-permeability transform, the log-weights did correlate with data mismatch when the standard IES method was used for data assimilation ($r=-0.416$) and when the hybrid IES method was used ($r=-0.647$).  In both cases, the quality of the data mismatch provided some information on the weighting that should be applied to a particle.
For the more highly nonlinear 2D porous flow example with non-monotonic log-permeability transform, the correlation between importance log-weight and data mismatch was perfect ($r=-0.813$) when the hybrid IES was used for data assimilation and the data mismatch could serve as a useful tool for eliminating samples with small weights. For the standard IES, however, the approximated weights are not accurate and the correlation between log-weight and data mismatch was correspondingly small ($r=-0.087$).  In this case, the data mismatch would not have provided a useful proxy for weighting.

\subsubsection{Effect of denoising weights on predictability}  

Unweighted posterior realizations generated by minimization of a stochastic cost function are often described as well history-matched, but differences in the quality of the match to data between some realizations and observations are too large in practice to be explained by observation error.  
To investigate the potential benefit of weighting the samples  and of different degrees of denoising we compute the accuracy of probabilistic predictions beyond the history matching period for data assimilation using the hybrid IES.  (We did not investigate optimal weighting for the standard IES as weighting was not useful for the non-monotonic log-permeability case.)  For this investigation, observations used in history matching end at $t=60$ and predictions are evaluated at $t=70$ for all 9 producers using the ``log score`` \cite{good:52,gneiting:07a}.  The Logarithmic score evaluates the probability of the outcome given a pdf empirically defined by the ensemble of predictions. The log score rewards both accuracy and sharpness of the forecasts. A higher log score signifies better probabilistic prediction:  
\begin{equation*}
\rm {LogS}(P,u) = - \log( p(u) )
\end{equation*}
where a Gaussian approximation of $\rm p$ has been used. 

\begin{figure}[htbp]
\begin{subfigure}{0.45\textwidth}
\includegraphics[width=0.95\textwidth]{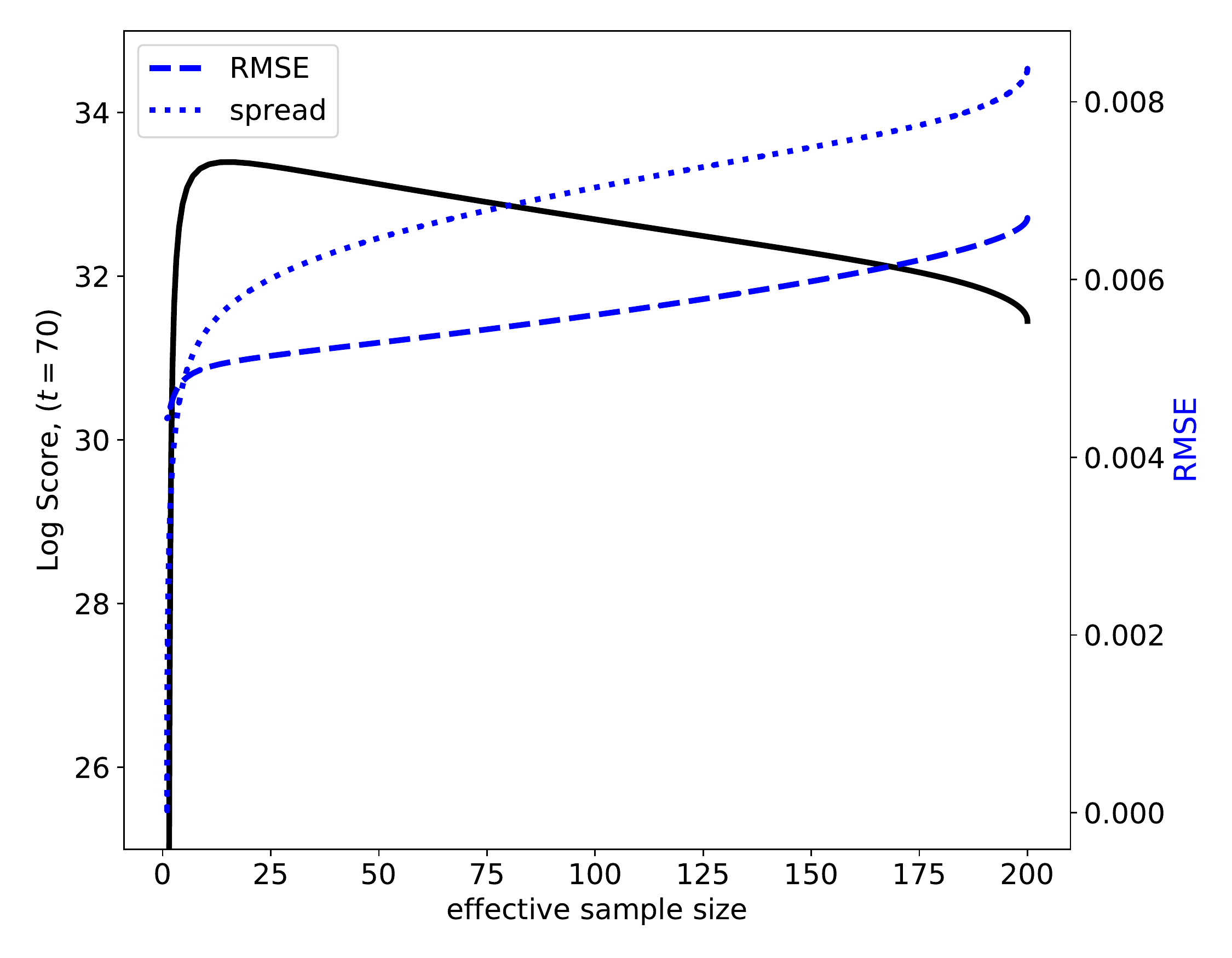}
\caption{monotonic}
\label{fig:optimal_efficiency_a}
\end{subfigure}
\hfill
\begin{subfigure}{0.45\textwidth}
\includegraphics[width=0.95\textwidth]{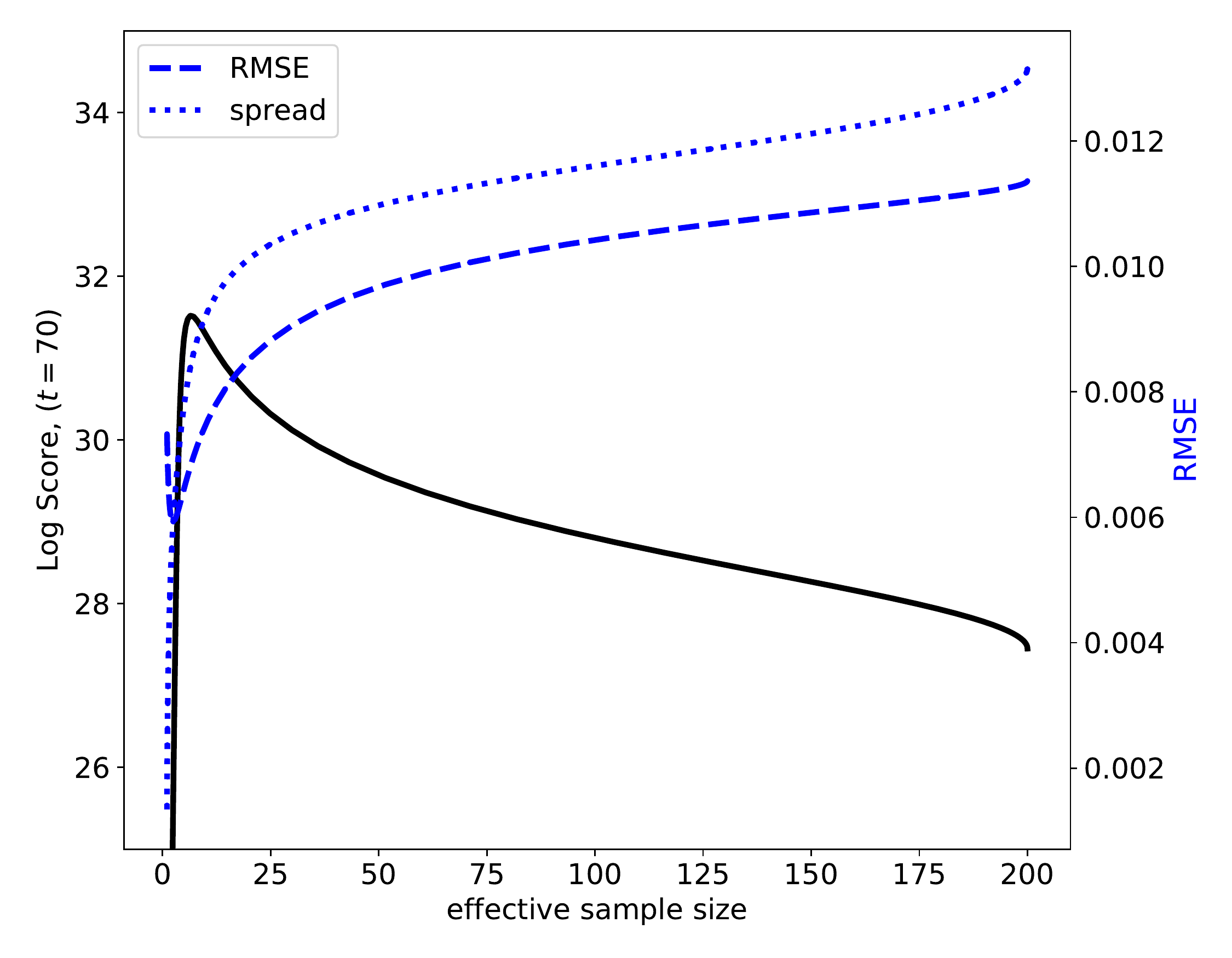}
\caption{non-monotonic}
\label{fig:optimal_efficiency_b}
\end{subfigure} 
\caption{Evaluation of optimal regularization of weights for forecast  predictability using hybrid IES.}
\label{fig:optimal_efficiency}
\end{figure}

We computed the effective sample size  and the log score of the forecasts   at $t= 70$  for all nine wells for a range of degrees of regularization of weights.  Regularization of log weights was accomplished by applying a power transformation with exponents between 0 and 1 to the computed weights.  The effective sample size (ess) is affected by the degree of regularization -- the ess is 200 if we use equal weighting (i.e.\ if we apply a power transformation with a very small exponent).  When the exponent is close to one, we end up using the weights as computed without denoising or regularization. 
Figure~\ref{fig:optimal_efficiency_a} shows predictability scores for a range of degrees of regularization using the hybrid iterative  smoother with the monotonic permeability transform. Figure~\ref{fig:optimal_efficiency_b} shows corresponding results for the non-monotonic permeability transform.  The solid black curves in both cases show the log score, which is somewhat small for both cases when the effective sample size is small, even though only the ``best'' realisations are used for the forecast.
The poor predictability for small effective sample size is a result of the small spread in the ensemble so that even small inaccuracy of the prediction is highly improbable. As the effective sample size increases, the predictability initially increases rapidly because of the increase in the spread, but when the exponent of the power transform is decreased sufficiently the predictability gradually decreases as more ``bad'' samples are added.
The impact of ``bad'' samples is smaller in the monotonic case than the non-monotonic case because the RMSE in the worst samples is smaller in the monotonic case.

\begin{figure}[htbp]  
\begin{tabular}{cccc}  
     & unweighted  &   weighted   &  denoised  \\
\raisebox{7ex}{\rotatebox{90}{monotonic}} &
\includegraphics[width=0.3\textwidth]{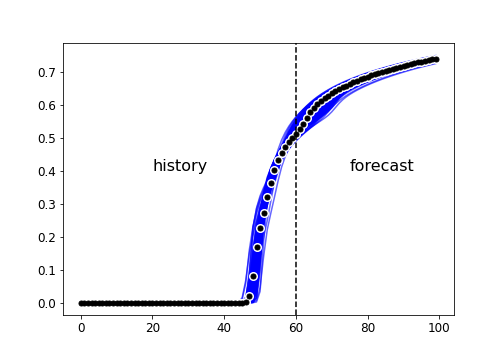} &
\includegraphics[width=0.3\textwidth]{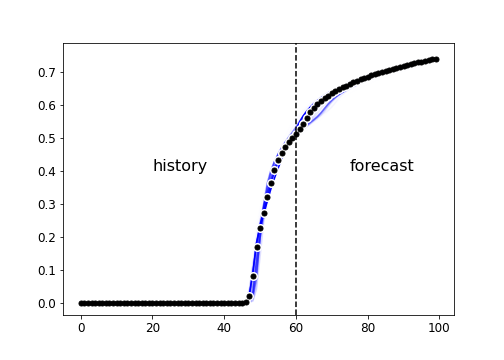} &
\includegraphics[width=0.3\textwidth]{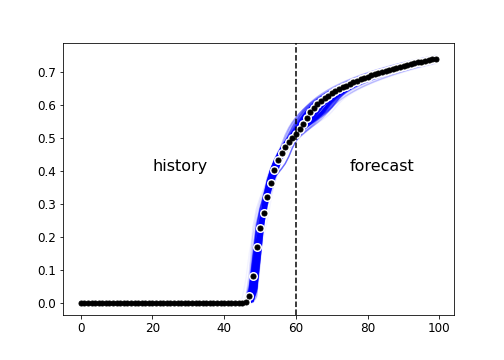} \\
\raisebox{4ex}{\rotatebox{90}{non-monotonic}} &
\includegraphics[width=0.3\textwidth]{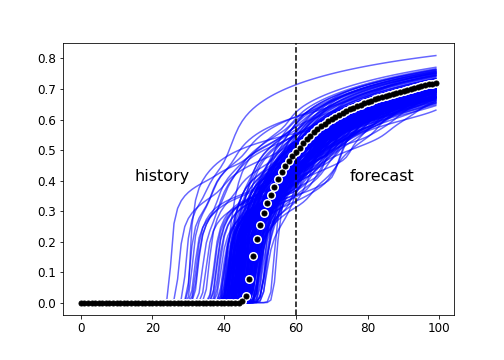} &
\includegraphics[width=0.3\textwidth]{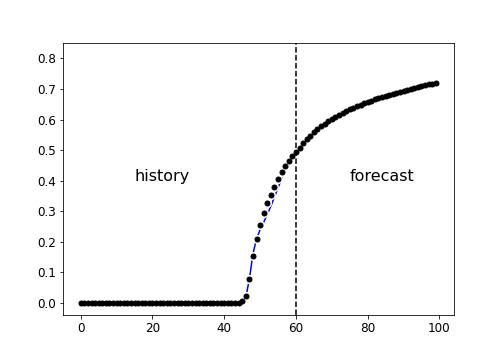} &
\includegraphics[width=0.3\textwidth]{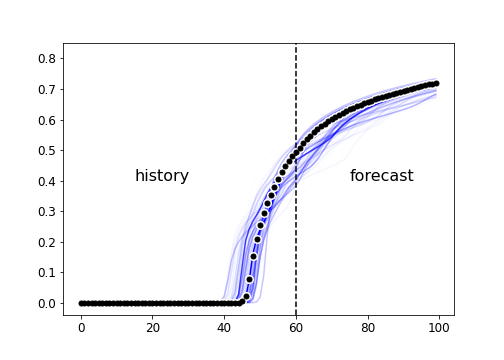}
\end{tabular}
\caption{The posterior distribution of forecasts conditioned to data to $t=60$ using hybrid IES method. The black dots show the observed data.}
\label{fig:forecast_weighting}
\end{figure}

Figure~\ref{fig:forecast_weighting} compares unweighted predictions to weighted predictions and denoised weighted predictions for one of the wells (Producer 4) in both 2D porous flow examples. 
For Producer 4,   the agreement between the forecast from the data-generating model and \emph{weighted} forecasts is nearly perfect, although the quality of the agreement at some other wells is less.  
Better forecast predictability as measured by the log score is obtained using denoised importance weights as described in Sec.~\ref{subsec:denosing}, although in both examples (monotonic and non-monotonic log permeability transforms) the correct level of denoising was difficult to determine.  

Although the effect of denoising is quantitatively different for the monotonic and non-monotonic permeability transforms, in both cases the best predictability is obtained when the weights are regularized such that the effect sample size is intermediate between the ess for unweighted samples and the ess for the weights computed using Eq.~\eqref{eq:wei}.


\section{Landscape of the posterior}

The efficiency of the  hybrid iterative smoother for sampling the posterior was relatively low in the flow example with the non-monotonic permeability transformation. For an an ensemble size of 200, the effective sample size after denoising was approximately 5.5. The small size makes probabilistic inference difficult -- even though the mean weighted forecasts  were generally accurate, the estimates of the uncertainty were often not.  In order to get an effective ensemble size of approximately 40 after weighting, it would be necessary to use an initial ensemble size of approximately 1600.  As the efficiency of the randomized maximum likelihood sampler using BFGS for minimization with the gradient computed from the adjoint system was similar to the efficiency of the hybrid IES for a similar problem \cite{ba:22}, it seems likely that the low efficiency is a result of the roughness of the posterior landscape rather than a problem with minimization. 

The efficiency of sampling algorithms depends strongly on the landscape of the pdf to be sampled and on the goal of the sampling. If the objective is simply to sample in the neighborhood of the maximum a posteriori point, then using exact gradients are not always beneficial, especially if the log posterior is characterized by multiple scales -- a smooth, long range feature that is approximately quadratic and shorter range fluctuations to the surface \cite{plechac:20}.  If the posterior pdf is characterized, however, by a small number of nearly equivalent modes, then ensemble methods may fail to converge \cite{oliver:18a,dunbar:22}.  In the numerical example with non-monotonic transformation of the log-permeability, the IES converged to the MAP, but failed to sample other local minima.  In order to clarify the behavior, we evaluated the fitness landscape in the neighborhood of the true data-generating model and over a larger region.

\begin{figure}[htbp]
\begin{subfigure}{0.47\textwidth}
\begin{tabular}{c}
\includegraphics[width=0.98\textwidth]{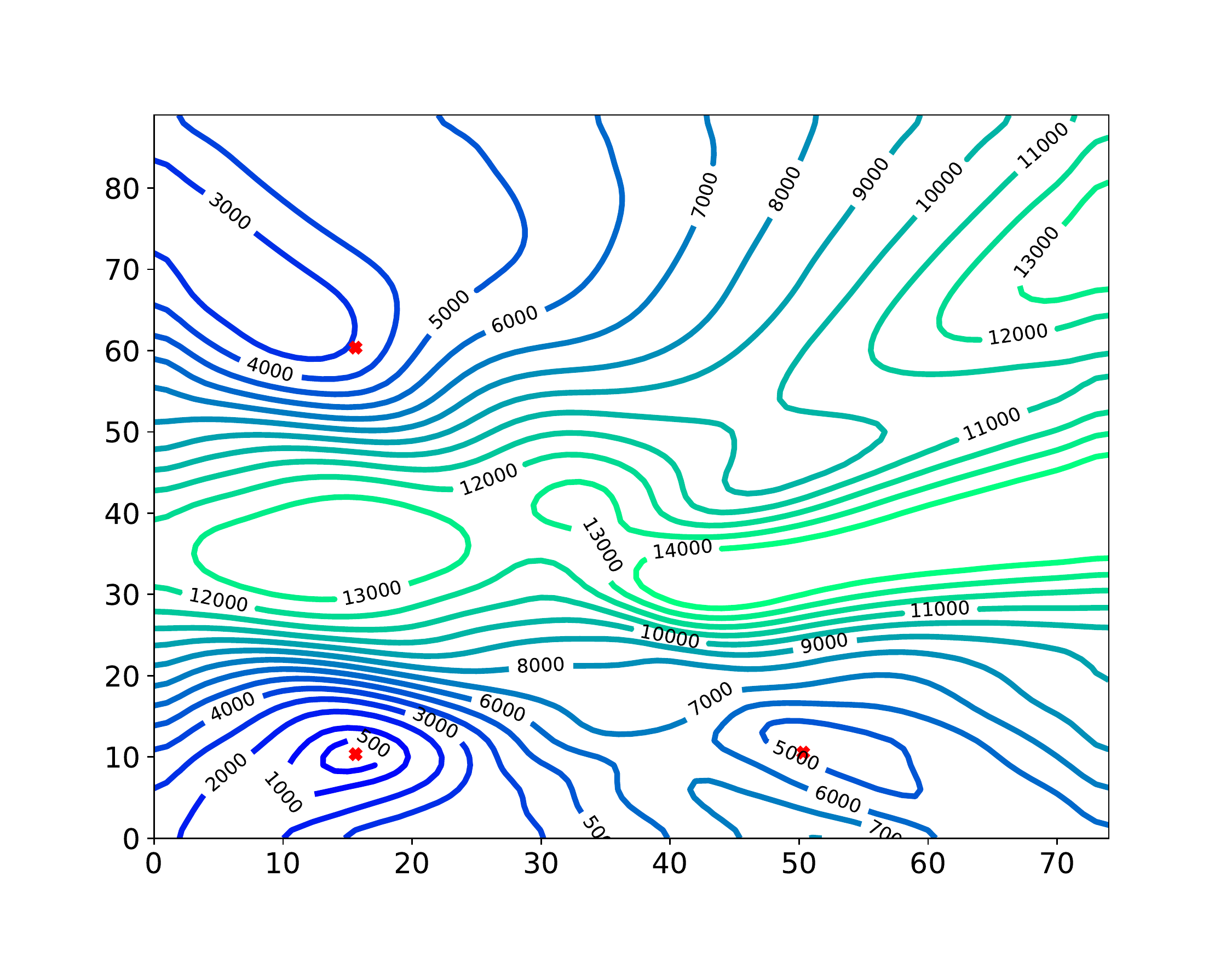}
\\
\includegraphics[width=0.98\textwidth]{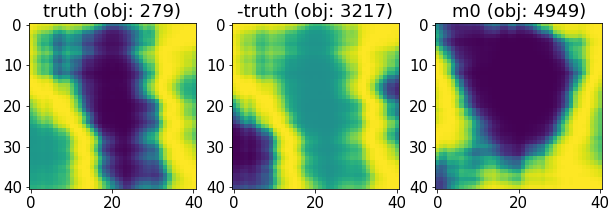}
\end{tabular}
\caption{Subspace generated by three very different realizations}
\label{fig:landscape_posterior_a}
\end{subfigure}
\hfill
\begin{subfigure}{0.47\textwidth}
\begin{tabular}{c}
\includegraphics[width=0.98\textwidth]{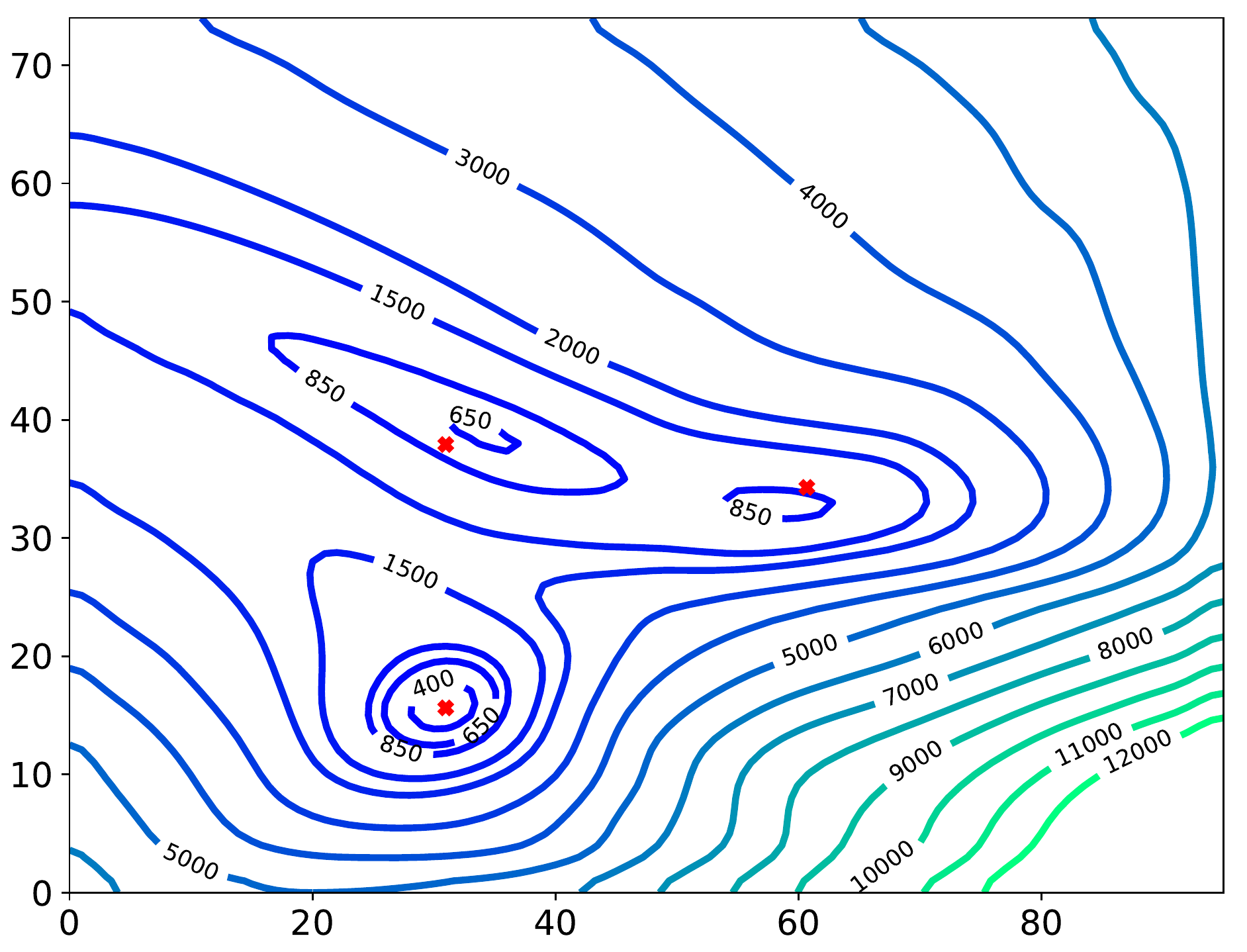}
\\
\includegraphics[width=0.98\textwidth]{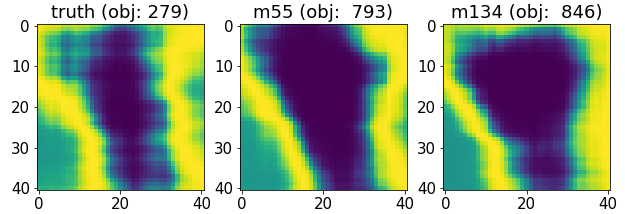}
\end{tabular}
\caption{Subspace generated by three similar realizations}
\label{fig:landscape_posterior_b}
\end{subfigure} 
\caption{The fitness landscape for the non-monotonic porous flow problem.}
\label{fig:landscape_posterior}
\end{figure}

The dimension of the model space is too large to visualize the landscape of the posterior directly. Instead, two illustrations of the fitness landscape for the non-monotonic log-permeability transform case were created by selecting three realizations of the model parameters to define a subspace of the  model space. From  three realizations, we constructed an orthonormal basis and evaluated the log-likelihood function on a grid containing the three realizations. In the first plot of the fitness landscape (Fig.~\ref{fig:landscape_posterior_a}) the subspace contains both the true model, $x\sbr{true}$, and the negative of the true model, $-x\sbr{true}$. (Both are equally probably before conditioning to the data.) A third realization with relatively low weight was included to provide an independent basis vector needed for the 2D subspace. In this case, there is a fairly large energy barrier separating the modes containing $x\sbr{true}$ and $-x\sbr{true}$ and a smaller barrier separating $x\sbr{true}$ from $x_0$. In Fig.~\ref{fig:landscape_posterior_b} the subspace contains the truth and two realizations with large posterior weights. Again, each realization appears to lie in a separate mode of the likelihood, although that cannot be verified without examining the surface in higher dimensions. In any case, the posterior landscape is complex and accurate gradients may be of limited usefulness if the goal is to locate the global minimum.


\section{Summary and conclusions}
\label{sec:conclusions}

Iterative, ensemble-based data assimilation methods for sampling the posterior distribution are based on minimisation of stochastic objective functions.
These methods of sampling are approximate when the mapping from parameters to observations is nonlinear. 
To correct the sampling, importance weighting can be used.  It is, however,  generally difficult or costly to compute the importance weights if derivatives are computed from the adjoint system. On the other hand, the cost is relatively low for ensemble-based methods which avoid the need for adjoints. We showed that standard products from hybrid iterative ensemble smoothers could be used to compute approximations to the importance weights.  The weights computed in this way are noisy -- largely because of low-rank stochastic approximations of derivatives. 
Denoising of the importance weights increased the effective sample size, decreased the RMSE in the estimate of the posterior model mean, and increased the predictability of future reservoir behavior. 

Although the IES method converged more  quickly than the hybrid IES in the numerical test problem with multimodal posterior, the posterior realizations from IES appear to be samples from a single mode. In some cases, the IES sampled from the mode with the highest probability so that while the uncertainty was underestimated, the fit to data was good. In other cases, however, the IES samples were centered on a mode with lower mass and the fit was less good. The posterior mean model for the multimodal problems using IES was very sensitive to the choice of minimization parameters.  Weighting was not effective in this case because the posteriori distributions of samples were not from the critical points of the stochastic cost function. We did not evaluate the possibility of combining multiple posterior ensembles from IES but it seems likely that the weighted results from a large number of ensembles would provide a better representation of the posterior.

Finally, we noted that the posterior landscape of the inverse problem for the 2D, 2-phase immiscible flow appears to be multimodal when the permeability field is generated from a transformation that creates channel-like features of high permeability in a low permeability background.  The characteristics of the posterior distribution have implications for the types of data assimilation methods that can be expected to provide reasonable assessment of the uncertainty. For this problem, it appears that it would be difficult to have confidence in an uncertainty quantification assessment from a standard iterative ensemble smoother, even if weighting was used. On the other hand, the hybrid iterative ensemble smoother was relatively inefficient as it sampled many local minima with low mass, but the results appear to be useful.

%
%
\appendix
\section{Nomenclature}
\begin{table*}[ht]
\begin{center}

\begin{tabular}{llll}
\multicolumn{1}{l}{List of notations}\\
\hline
$N_x$ &number of model parameters& ${\rm N}(\cdot,\cdot)$ & Gaussian distribution\\
$N_d$&dimension of observational space& $C_m$& covariance matrix of Gaussian prior for $m$\\
$N_e$&number of samples&$C_d$& covariance matrix of Gaussian observation error  \\
$N_m$ &number of intermediate variable& 
$C_x$&covariance matrix of Gaussian prior for $x$\\
$N_{\text{eff}}$&effective sample size&$w$&weights\\
$m$&intermediate variable&$\theta$&hyparameter or other reservoir properties\\
$n_x$ and $n_y$ &number of grid nodes &$h_x$ and $h_y$ &mesh size\\
$d^o$ & observations&$x^{\text{pr}}$& mean of Gaussian prior  \\
$x_l^i$ & $i$th sample for the $l$th iteration&$\epsilon$& observation error\\
$\mathcal{B}$ & generic forward model operator&$l$ & iteration index\\
$g$ & observation operator & $\pi_D(d^{\rm o})$&normalisation constant\\
$O(x)$&negative log likelihood function&$\pi_X(x|d^{\rm o})$& posterior distribution\\
$\pi_{X\Delta} (x,\delta)$& target distribution&$q_{X'\Delta'} (x',\delta')$& proposal distribution of $(x',\delta')$\\
 $p_{X\Delta} (x,\delta)$& proposal distribution of $(x,\delta)$& $\kappa _{ri}$&relative permeability of phase i\\
 $J$&Jacobian determinant&$\mu _{i}$& viscosity of phase i\\
 $x$ and $\delta$& samples of target distribution&$x'$ and $\delta'$ & samples of Gaussian $q_{X'\Delta'} (m',\delta')$\\
 $Z$ & Composite observation operator &$x^*$ and $\delta^*$ & prior samples\\
$Z_{u,\theta}$& differential operator of $Z$ &$V$, $\eta(m)$ &auxiliary variables\\
$u(x)$& model state &$p(x,t)$ & pressure \\
$s(x,t)$& saturation&$\Omega$& spatial domain\\
$n(x')$ &total number of critical points&$G$& differential operator of $g$ wrt $x$\\
$G_m$ & differential operator of $g$ wrt $m$ & $M_x$& differential operator of $m$ wrt $x$\\
$L$ & square root of $C_M$ &$\lambda_l$& $l$th regularization parameter\\
$I_N$ &$N$-dimensional identity matrix &$1_{N}$& $N$-dimensional one-vector\\
$\Delta x$ & ensemble deviation from parameter mean &$\Delta d$ &ensemble deviation from data mean\\
$\delta x_l$ & increment for the $l$th iteration &$\Delta m$ &ensemble deviation from mean of $m$ \\
$\check{C} $ & circulant matrix augmented by $C_x$&$\check{c} $ & first column of $\check{C}$\\
$P(w|w^o) $ & posterior distribution of weights&$w^o $ &observation of weights\\
$P(w)$ & prior of weights&$P(w^o|w)$ & likelihood function of weights\\
$\nu$ &freedom of Chi-square prior for $w$& $\sigma_{pr} $ &standard deviation of prior for $w$\\
$\sigma_o $ & standard deviation of $w^o$ & $C$ 
&covariance operator\\
$\rho $ & correlation length of $C$ & $\sigma$ & standard deviation of $C$\\
$\rm p(u)$ & predictive distribution of state $u$ & $f$ & nonlinear transform from $x$ to $m$\\
bold letter & vector or matrix of corresponding letter & unbold letter &analytic representation or scalar\\
\hline
\end{tabular}
\end{center}
\caption{Notation used throughout the manuscript.}
\label{tab:Notations}
\end{table*}
%
\bibliographystyle{siam}


\begin{thebibliography}{10}

\bibitem{aarnes:07}
{\sc J.~E. Aarnes, T.~Gimse, and K.-A. Lie}, {\em An introduction to the
  numerics of flow in porous media using {M}atlab}, in Geometric Modelling,
  Numerical Simulation, and Optimization, G.~Hasle, K.~Lie, and E.~Quak, eds.,
  Springer, 2007, pp.~265--306.

\bibitem{acerbi:20}
{\sc L.~Acerbi}, {\em Variational {B}ayesian {M}onte {C}arlo with noisy
  likelihoods}, in Advances in Neural Information Processing Systems,
  H.~Larochelle, M.~Ranzato, R.~Hadsell, M.~Balcan, and H.~Lin, eds., vol.~33,
  Curran Associates, Inc., 2020, pp.~8211--8222.

\bibitem{akyildiz:17}
{\sc O.~D. Akyildiz, I.~P. Marino, and J.~M{\'\i}guez}, {\em Adaptive noisy
  importance sampling for stochastic optimization}, in 2017 IEEE 7th
  International Workshop on Computational Advances in Multi-Sensor Adaptive
  Processing (CAMSAP), IEEE, 2017, pp.~1--5.

\bibitem{alquier:16}
{\sc P.~Alquier, N.~Friel, R.~Everitt, and A.~Boland}, {\em Noisy {M}onte
  {C}arlo: {C}onvergence of {M}arkov chains with approximate transition
  kernels}, Statistics and Computing, 26 (2016), pp.~29--47.

\bibitem{ba:22}
{\sc Y.~Ba, J.~de~Wiljes, D.~S. Oliver, and S.~Reich}, {\em Randomized maximum
  likelihood based posterior sampling}, Computat. Geosci., 26 (2022),
  pp.~217--239.

\bibitem{ba:21}
{\sc Y.~Ba and L.~Jiang}, {\em A two-stage variable-separation {K}alman filter
  for data assimilation}, Journal of Computational Physics, 434 (2021),
  p.~110244.

\bibitem{bardsley:14}
{\sc J.~Bardsley, A.~Solonen, H.~Haario, and M.~Laine}, {\em
  Randomize-{T}hen-{O}ptimize: A method for sampling from posterior
  distributions in nonlinear inverse problems}, SIAM Journal on Scientific
  Computing, 36 (2014), pp.~A1895--A1910.

\bibitem{bardsley:20}
{\sc J.~M. Bardsley, T.~Cui, Y.~M. Marzouk, and Z.~Wang}, {\em Scalable
  optimization-based sampling on function space}, SIAM Journal on Scientific
  Computing, 42 (2020), pp.~A1317--A1347.

\bibitem{chen:12}
{\sc Y.~Chen and D.~S. Oliver}, {\em Ensemble randomized maximum likelihood
  method as an iterative ensemble smoother}, Mathematical Geosciences, 44
  (2012), pp.~1--26.

\bibitem{chen:13}
{\sc Y.~Chen and D.~S. Oliver}, {\em Levenberg-{M}arquardt forms of the
  iterative ensemble smoother for efficient history matching and uncertainty
  quantification}, Computational Geosciences, 17 (2013), pp.~689--703.

\bibitem{chen:17}
{\sc Y.~Chen and D.~S. Oliver}, {\em Localization and regularization for
  iterative ensemble smoothers}, Computational Geosciences, 21 (2017),
  pp.~13--30.

\bibitem{dietrich:97}
{\sc C.~R. Dietrich and G.~N. Newsam}, {\em Fast and exact simulation of
  stationary {G}aussian processes through circulant embedding of the covariance
  matrix}, SIAM Journal on Scientific Computing, 18 (1997), pp.~1088--1107.

\bibitem{dunbar:22}
{\sc O.~R.~A. Dunbar, A.~B. Duncan, A.~M. Stuart, and M.-T. Wolfram}, {\em
  Ensemble inference methods for models with noisy and expensive likelihoods},
  SIAM Journal on Applied Dynamical Systems, 21 (2022), pp.~1539--1572.

\bibitem{emerick:13}
{\sc A.~A. Emerick and A.~C. Reynolds}, {\em Ensemble smoother with multiple
  data assimilation}, Computers \& Geosciences, 55 (2013), pp.~3--15.

\bibitem{evensen:94}
{\sc G.~Evensen}, {\em Sequential data assimilation with a nonlinear
  quasi-geostrophic model using {M}onte {C}arlo methods to forecast error
  statistics}, Journal of Geophysical Research: Oceans, 99 (1994),
  pp.~10143--10162.

\bibitem{gneiting:07a}
{\sc T.~Gneiting and A.~E. Raftery}, {\em Strictly proper scoring rules,
  prediction, and estimation}, Journal of the American Statistical Association,
  102 (2007), pp.~359--378.

\bibitem{good:52}
{\sc I.~J. Good}, {\em Rational decisions}, J. R. Stat. Soc. Ser. B-Stat.
  Methodol., 14 (1952), pp.~107--114.

\bibitem{kitanidis:95}
{\sc P.~K. Kitanidis}, {\em Quasi-linear geostatistical theory for inversing},
  Water Resour. Res., 31 (1995), pp.~2411--2419.

\bibitem{martin:12}
{\sc J.~Martin, L.~Wilcox, C.~Burstedde, and O.~Ghattas}, {\em A stochastic
  {N}ewton {MCMC} method for large-scale statistical inverse problems with
  application to seismic inversion}, SIAM Journal on Scientific Computing, 34
  (2012), pp.~A1460--A1487.

\bibitem{maschio:14}
{\sc C.~Maschio and D.~J. Schiozer}, {\em Bayesian history matching using
  artificial neural network and {M}arkov {C}hain {M}onte {C}arlo}, Journal of
  Petroleum Science and Engineering, 123 (2014), pp.~62--71.
\newblock Neural network applications to reservoirs: {P}hysics-based models and
  data models.

\bibitem{mohamed:12}
{\sc L.~Mohamed, B.~Calderhead, M.~Filippone, M.~Christie, and M.~Girolami},
  {\em Population {MCMC} methods for history matching and uncertainty
  quantification}, Computational Geosciences, 16 (2012), pp.~423--436.

\bibitem{oliver:17}
{\sc D.~S. Oliver}, {\em Metropolized randomized maximum likelihood for
  improved sampling from multimodal distributions}, SIAM/ASA Journal on
  Uncertainty Quantification, 5 (2017), pp.~259--277.

\bibitem{oliver:22a}
{\sc D.~S. Oliver}, {\em Hybrid iterative ensemble smoother for history
  matching of hierarchical models}, Mathematical Geosciences, 54 (2022),
  pp.~1289--1313.

\bibitem{oliver:11}
{\sc D.~S. Oliver and Y.~Chen}, {\em Recent progress on reservoir history
  matching: a review}, Computational Geosciences, 15 (2011), pp.~185--221.

\bibitem{oliver:18a}
{\sc D.~S. Oliver and Y.~Chen}, {\em Data assimilation in truncated
  plurigaussian models: impact of the truncation map}, Mathematical
  Geosciences, 50 (2018), pp.~867--893.

\bibitem{oliver:97}
{\sc D.~S. Oliver, L.~B. Cunha, and A.~C. Reynolds}, {\em Markov chain {Monte}
  {Carlo} methods for conditioning a permeability field to pressure data},
  Mathematical Geology, 29 (1997), pp.~61--91.

\bibitem{oliver:96e}
{\sc D.~S. Oliver, N.~He, and A.~C. Reynolds}, {\em Conditioning permeability
  fields to pressure data}, in Proceedings of the European Conference on the
  Mathematics of Oil Recovery, V, 1996, pp.~1--11.

\bibitem{oliver:11b}
{\sc D.~S. Oliver, Y.~Zhang, H.~A. Phale, and Y.~Chen}, {\em Distributed
  parameter and state estimation in petroleum reservoirs}, Computers \& Fluids,
  46 (2011), pp.~70--77.

\bibitem{papaspiliopoulos:07}
{\sc O.~Papaspiliopoulos, G.~O. Roberts, and M.~Sk{\"o}ld}, {\em A general
  framework for the parametrization of hierarchical models}, Statist. Sci., 22
  (2007), pp.~59--73.

\bibitem{plechac:20}
{\sc P.~Plech\'{a}\v{c} and G.~Simpson}, {\em Sampling from rough energy
  landscapes}, Communications in Mathematical Siences, 18 (2020),
  pp.~2271--2303.

\bibitem{reich:11}
{\sc S.~Reich}, {\em A dynamical systems framework for intermittent data
  assimilation}, BIT Numerical Mathematics, 51 (2011), pp.~235--249.

\bibitem{tavassoli:05}
{\sc Z.~Tavassoli, J.~N. Carter, and P.~R. King}, {\em An analysis of history
  matching errors}, Comput. Geosci., 9 (2005), pp.~99--123.

\bibitem{vanLeeuwen:19}
{\sc P.~J. van Leeuwen, H.~R. K{\"u}nsch, L.~Nerger, R.~Potthast, and
  S.~Reich}, {\em Particle filters for high-dimensional geoscience
  applications: A review}, Quarterly Journal of the Royal Meteorological
  Society, 145 (2019), pp.~2335--2365.

\bibitem{zhang:03b}
{\sc F.~Zhang, A.~C. Reynolds, and D.~S. Oliver}, {\em The impact of upscaling
  errors on conditioning a stochastic channel to pressure data}, SPE Journal, 8
  (2003), pp.~13--21.

\bibitem{zimmerman:89}
{\sc D.~L. Zimmerman}, {\em Computationally exploitable structure of covariance
  matrices and generalized convariance matrices in spatial models}, Journal of
  Statistical Computation and Simulation, 32 (1989), pp.~1--15.

\end{thebibliography}

\end{document}